\documentclass[prb,twocolumn,superscriptaddress,nofootinbib,floatfix]{revtex4-2}
\usepackage{soul}
\usepackage{amsfonts}
\usepackage{graphicx}
\usepackage{bm}
\usepackage{amssymb}
\usepackage{amsmath}
\usepackage{amstext}
\usepackage{latexsym}
\usepackage{braket}
\usepackage{enumitem}
\usepackage[normalem]{ulem}
\usepackage[usenames,dvipsnames]{xcolor}
\usepackage[colorlinks=true,citecolor=Cerulean,linkcolor=RubineRed,urlcolor=Cerulean]{hyperref}
\definecolor{crimson}{RGB}{220,20,60}
\newcommand\ba{\begin{eqnarray}}
\newcommand\ea{\end{eqnarray}}
\newcommand\be{\begin{equation}}
\newcommand\ee{\end{equation}}

\usepackage{braket}
\usepackage{float}

\newcommand\bka[1]{\textcolor{magenta}{#1}}

\usepackage{verbatim}
\usepackage{graphicx}
\begin{document}
\author{Tamizhselvan S}
%\email{}
 \affiliation{Department of Physics, Indian Institute of Technology Palakkad, Palakkad, Kerala 678623, India}
\author{Manju C}
%\email{}
\affiliation{Department of Physics, Indian Institute of Technology Palakkad, Palakkad, Kerala 678623, India}
 
\author{Bijay Kumar Agarwalla}
%\email{bijay@iiserpune.ac.in}
\affiliation{Department of Physics, Indian Institute of Science Education and Research Pune, Dr. Homi Bhabha Road, Ward No. 8, NCL Colony, Pashan, Pune, Maharashtra 411008, India
}

\author{Uma Divakaran}
%\email{uma@iitpkd.ac.in}
 \affiliation{Department of Physics, Indian Institute of Technology Palakkad, Palakkad, Kerala 678623, India}

\title{Symmetry structure dependent diagnostic of the Quantum Mpemba Effect}

%TITLE SUGGESTION
% 1. Symmetry Structure and Diagnostic Dependence of the Quantum Mpemba Effect
% 2. Symmetry Structure Controls the Diagnostic Dependence of the Quantum Mpemba Effect
% 3. Symmetry

\newcommand{\shortabstract}{
Understanding symmetry restoration in isolated quantum many-body systems is an important problem in nonequilibrium many-body quantum physics. Recent studies have shown that the quantum Mpemba effect can be characterized through entanglement asymmetry, where states with stronger initial symmetry breaking restore symmetry faster. However, it remains unclear whether conventional energy-based measures, such as the trace distance, capture the same phenomenon. We investigate this question in closed spin-$1/2$ quantum systems with different symmetries by analyzing the dynamics of symmetry-breaking initial states. Combining numerical simulations with an analytical decomposition of the trace distance into symmetry-coherence and residual contributions, we identify the conditions under which trace distance tracks entanglement asymmetry and reproduces the Mpemba--like behavior observed in it. For charge symmetry, the residual contribution is negligible, making the trace distance effectively governed by symmetry-sector coherences. In contrast, for permutation symmetry, a significant residual contribution leads to qualitatively different relaxation dynamics. Our results establish when conventional energy-based diagnostics reliably capture symmetry-restoration dynamics and clarify the  distinct physical information encoded by entanglement asymmetry and trace distance.
}

\newcommand{\longabstract}{
{\sout{Understanding symmetry-breaking and its restoration, along with thermalization in isolated quantum many-body systems is central to nonequilibrium quantum physics. The idea of the Mpemba effect, where a system prepared farther from equilibrium relaxes faster than the one prepared closer to it, has been studied in classical systems and has recently been extended to quantum systems.} In quantum systems, entanglement asymmetry has emerged as a faithful probe of Mpemba-like effect, where the relaxation dynamics is characterized in terms of symmetry restoration. In particular, an initial state with a higher degree of symmetry breaking relaxes more rapidly under time evolution generated by a Hamiltonian that preserves the corresponding symmetry, resembling Mpemba effect in classical systems where a system prepared farther from equilibrium relaxes faster than the one prepared closer to it. On the other hand, it is not clear the extent to which the conventional purely energy-based relaxation measures such as trace distance can successfully capture the Mpemba effect. To probe this further, we study the dynamics of closed spin-1/2 Hamiltonians showing different symmetries,
and study the non-equilibrium dynamics by choosing the initial states which break the particular symmetry studied. 
Using numerical simulations together with an analytical decomposition of the trace distance into symmetry-coherence and residual contributions, we identify the conditions under which the two diagnostics, namely, entanglement asymmetry and trace distance, exhibit equivalent relaxation behavior. Focusing on charge symmetry and permutation symmetry, we show that for charge symmetry, the residual contribution is very weak \textcolor{red}{for the Hamiltonians studied,} causing the trace distance to be governed by the decay of symmetry-sector coherences, resembling the Mpemba--like relaxation shown by the entanglement asymmetry. In contrast, for permutation symmetry, this residual contribution evolves nontrivially in time, leading to qualitatively different behavior of these two diagnostics. These results establish the symmetry--based conditions under which energy based diagnostics can serve as a clean indicator of Mpemba--like relaxation, and clarify the distinct physical information denoted by entanglement asymmetry and conventional relaxation measures in isolated quantum systems.}}

\date{\today}

\begin{abstract}
\shortabstract

%**Abstract**

\end{abstract}

%\begin{abstract}
%\bka{Understanding Mpemba effect in the relaxation dynamics of symmetry restoration while starting with symmetry broken initial state has is central to nonequilibrium quantum physics. In quantum systems, entanglement asymmetry has emerged as a faithful probe of Mpemba-like effect, }


\maketitle
%%%%%%%%%%%%%%%%%%%%%%%%%%%%%%%%%%%%%%%%%%%%%%%%%%%%%%%%%%%%%%%%%%%%%

%\textcolor{blue}{Replace $t_Q$ by $\tau$}\\

\section{Introduction}
Mpemba effect \cite{Mpemba_1969} refers to the counterintuitive phenomenon in which a system prepared farther from equilibrium relaxes towards the equilibrium faster than the one prepared closer to it. Originally discovered and studied in classical macroscopic systems like water and other aqueous solutions \cite{Mpemba_1969,Freeman1979_water,walker1977_water,auerbach1995supercooling,wojciechowski1988_aqueous_ME}, this anomalous relaxation has attracted considerable attention across a varied range of classical settings \cite{lasanta2017_ME,Klinch_ME_2019,maxwellgas2020_ME,Bechhoefer2021,mompo2021_ME,kumar2022_ME,spinsystem2024_ME,malhotra2024_ME,thermomajorization2025_ME}. 
%This idea is later extended to quantum systems as \emph{quantum Mpemba effect} (\rm{QME}), where it was \bka{NOT SURE IF OPEN SYSTEM WAS FIRST} \tam{I have got this info from a review article on "Quantum Mpemba Effects" by Ares,Calabrese,Murciano} first observed in open quantum systems \cite{carollo2021_openQMEorg} but subsequently studied both in open quantum systems  \cite{nava2019_openQMEorg,nava2024_openQME,wang2024_openQME,medina2025_openQME,qian2025_openQME,bka2026_openQME} and closed quantum systems undergoing unitary dynamics \cite{liu2024_closedQME,Chalas2024_closedQME,guo2025_closedQME,Calabrese2026_closedQME}.\bka{Maybe we can diilute as: }
The idea is later extended to quantum systems as \emph{quantum Mpemba effect} (\rm{QME}), where it was studied in the context of open systems \cite{carollo2021_openQMEorg, nava2019_openQMEorg,nava2024_openQME,wang2024_openQME,medina2025_openQME,qian2025_openQME,bka2026_openQME} as well as closed quantum systems undergoing unitary dynamics \cite{liu2024_closedQME,Chalas2024_closedQME,guo2025_closedQME,Calabrese2026_closedQME}. The \rm{QME} has been investigated theoretically in various spin models  \cite{microorigin2024_QME,Murciano2024_XYchain,Wei2026_spinQME,ganguly2026}, fermionic \cite{yamashika2_fermionicQME,yamashika2024_fermionicQME,Ares2025_FreeFermionic} and bosonic systems \cite{longhi2024_bosonicQME,longhi2025_mpembaeffectsuper,yamashika2025_bosonicQME}, quantum random circuits \cite{turkeshi2025_randcirQME,foligno2025_randcirQME,liHan2026_QME_U1} etc., and has also been experimentally observed \cite{Hofstetter2018,Joshi_QME_2024,chatterjee2025_expQME,xu2026_QME}, signifying its relevance in practical implications of quantum systems for  applications like gaining cooling advantage \cite{schnepper2025_appliQME}, faster entanglement generation \cite{bao2026}. In isolated quantum many-body systems, an analogue of the Mpemba effect has been identified in the context of symmetry-breaking dynamics, where states with stronger initial symmetry breaking restore the underlying symmetry more rapidly \cite{bertini2024_symQME,rylands2024_symQME,liu2025_symQME,giulio2025_symQME}. Consequently, understanding symmetry restoration dynamics has turned out to be a central topic in the study of non-equilibrium quantum many-body systems.

%A particularly striking example of anomalous relaxation is the Mpemba effect \cite{Mpemba_1969}, which generally refers to the phenomenon that a system prepared farther from equilibrium  reaches equilibrium faster. This counterintuitive anomalous behaviour sparked significant interest in studying this phenomenon in various classical systems \cite{Mpemba_1969,Klinch_ME_2019} and recently in quantum systems \underline{(cite)}}

%Understanding symmetry restoration dynamics in many-body isolated quantum systems has become a topic of significant recent 
%interests. 

In this context, entanglement asymmetry has emerged as an effective diagnostic of the quantum Mpemba effect \cite{Ares2023_EAoriginal,Ares2023_lack,Ares_2025,ares2025_EAQME}. However for certain symmetries like $U(1)$ charge conservation, some conventional measures of relaxation, based on the distance from equilibrium, such as the trace distance, Frobenius distance \cite{noGlobal}, relative entropy between reduced and stationary state \cite{simpleprobe2025_QME}, have also been found to qualitatively reproduce Mpemba--like relaxation of symmetry restoration dynamics, while at the same time providing a formulation of anomalous relaxation that remains meaningful even in systems without a global symmetry \cite{noGlobal}. This raises a natural question: 
%under what conditions of the symmetry considered, does the relaxation of symmetry--breaking coherences control the entire approach of a subsystem toward stationarity?(\bka{I didn't exactly follow this question? Can we rewrite)}
under what conditions, the approach of a subsystem towards a stationary state resemble the symmetry-breaking dynamics, thus showing Mpemba like relaxation?

In this work, we investigate the above question by studying  a closed spin-$1/2$ quantum system that exhibits two different symmetries for different combinations of the parameters, namely, charge symmetry, and permutation symmetry.  When identical spins are subjected to all-to-all interactions showing permutation symmetry, it constraints the dynamics and allows the exponentially large many-body Hilbert space to be replaced by a much smaller collective--spin description \cite{Kumari2022eigenstate,pannier2025_longrange}. At the same time, evolution of a permutation symmetry broken initial state with a Hamiltonian having permutation symmetry would not be restricted within the collective spin subspace. In particular, nonuniform spin configurations may be represented as several internally permutation-invariant parts, each with different orientations. Such inhomogeneous initial states populate different total--spin sectors when written in the total spin basis, and generate coherences between them. Similarly, charge symmetry--breaking dynamics can also be studied with this same collective spin system setting by tuning the system parameters, and a charge symmetry--breaking initial state which has coherences between different charge sectors. Here, we analyze the dynamics of both the entanglement asymmetry and the trace distance starting with these different class of symmetry-breaking initial states.

Using numerical simulations together with an analytical decomposition of the trace distance into symmetry-coherence and residual contributions, we identify when entanglement asymmetry and trace distance are expected to display similar relaxation. For charge symmetry, the residual contribution vanishes for collective spin systems. The trace distance is therefore mainly governed by the decay of symmetry-sector coherences and follows the Mpemba--like behavior seen in the entanglement asymmetry. In contrast, for permutation symmetry, the residual contribution remains dynamically important and changes significantly with time, causing the two diagnostics to evolve differently.  Our results establish when a conventional energy based diagnostics reliably capture symmetry restoration dynamics and indicates the distinct physical information denoted by entanglement asymmetry and trace distance.

This paper is organized as follows. In Sec.~(\ref{sec:model_symmetry_diagnostics}), we introduce the collective spin Hamiltonians, discuss the different symmetries under our consideration, define the symmetry-breaking initial states, and present the two relaxation diagnostics. We numerically compare different measures for the two symmetries considered in order to probe the Mpemba-like relaxation in Sec.~(\ref{sec:numerical_observation}). In Sec.~(\ref{sec:analytical}), we analytically explain the different behavior of diagnostics for the two different symmetry classes. We  summarize our results and discuss the broader implications of the diagnostic dependence of QME in Sec.~(\ref{sec:summary}). Additional discussions and calculations along with results for some generic cases are provided in the Appendices.

%%%%%%%%%%%%%%%%%%%%%%%%%%%%%%%%%%%%%%%%%%%%%%%%%%%%%%%%%%%%%%%%%%%%%
\newcommand{\Tr}{\operatorname{Tr}}
\newcommand{\rA}{\hat{\rho}_A}
\newcommand{\rDE}{\hat{\rho}_{\rm DE}}
\newcommand{\rADE}{\hat{\rho}^{A}_{\rm DE}}
\newcommand{\Dtd}{\Delta(\hat \rho_A)}
\newcommand{\EA}{\Delta S_A}
\newcommand{\Wop}{\hat{W}_A}
\newcommand{\dop}{\hat{\delta}_A}
\newcommand{\Q}{\hat{Q}}
\newcommand{\QA}{\hat{Q}_A}
\newcommand{\QB}{\hat{Q}_B}
\newcommand{\J}{\hat{\vec{J}}}
\newcommand{\JA}{\hat{\vec{J}}_A}
\newcommand{\JB}{\hat{\vec{J}}_B}
\newcommand{\Jtwo}{\hat{J}^2}
\newcommand{\JAtwo}{\hat{J}_A^2}
\newcommand{\JBtwo}{\hat{J}_B^2}
\newcommand{\id}{\hat{\mathbb{I}}}
\newcommand{\normone}[1]{\left\|#1\right\|_1}
\newcommand{\normtwo}[1]{\left\|#1\right\|_2}

\newcommand{\rQA}{\hat{\rho}_{A,\hat{Q}_A}}
\newcommand{\rJAtwo}{\hat{\rho}_{A,\hat{J}_A^2}}
\newcommand{\rACA}{\hat \rho_{A,\hat{\mathcal{C}}_A}}

%%%%%%%%%%%%%%%%%%%%%%%%%%%%%%%%%%%%%%%%%%%%%%%%%%%%%%%%%%%%%%%%%%%%%
% Main paper body after Introduction
%%%%%%%%%%%%%%%%%%%%%%%%%%%%%%%%%%%%%%%%%%%%%%%%%%%%%%%%%%%%%%%%%%%%%

\section{Model, symmetries, and diagnostics}
\label{sec:model_symmetry_diagnostics}

In this section, we define the collective Hamiltonian used to study both charge and permutational symmetry breaking dynamics, highlighting the differences between these symmetry structures, along with defining different symmetry-broken initial states which are used for studying the dynamics of a particular broken symmetry. We also define the diagnostic tools studied in this work to probe QME in these symmetry-breaking dynamics.

\subsection{Collective spin Hamiltonian}
\label{subsec:collective_hamiltonian}

We consider a system of $N$ spin-1/2 particles where each spin interacts with every other spin equally, so that the total Hamiltonian can be written using the collective angular momentum operators $\hat J_{\mu}$ as follows:
\begin{equation}
    \hat{H} =    \alpha_1 \hat{J}_x^2+
    \alpha_2 \hat{J}_y^2+
    p\hat{J}_z .
    \label{eq:H_general}
\end{equation}
Here, 
\( \hat{J}_{\mu}
    =
    \frac{1}{2}
    \sum_{i=1}^{N}
    \hat{\sigma}_{i}^{\mu} \;; \; \mu = x,y,z
\)
and \(\hat{\sigma}_{i}^{\mu}\) is the Pauli-\(\mu\) operator acting on the \(i\)-th spin \cite{PhysRevE.97.052209,PhysRevE.70.016217}. 
This Hamiltonian is useful since it allows us to study both charge symmetry and permutational symmetry within a common dynamical framework as we discuss below.

The Hamiltonian  in Eq.~\eqref{eq:H_general} commutes with the total angular momentum operator, i.e., \([\hat H,\Jtwo]=0\) so that $\hat{J}^2$ is a constant of motion. The above Hamiltonian shows permutation symmetry for any value of \(\alpha_1\,,\,\alpha_2\,,\text{and}\,\,p\) since each term of the Hamiltonian consists of  collective operators 
%The spin operator \(\Jtwo\) acts as a probe of permutation symmetry %because, for a
%system of \(N\) spin-\(\frac12\) particles, the fully permutation-symmetric
%subspace coincides with the maximum total-spin sector \(j=N/2\).
%Any finite weight in lower total-spin sectors \(j<N/2\) indicates leakage out of
%the fully symmetric subspace and hence signals permutation symmetry-breaking 
\cite{Ieminietal,manju2025disordering,manju_lakshmi2025chaos}.
The relation between $\Jtwo$ operator and permutation symmetry is discussed in details in Appendix \ref{apdx:J2_permutation}.

An additional symmetry, namely the charge symmetry, appears when \(\alpha_1=\alpha_2=\alpha\). In this case, 
\begin{equation}
    \hat{H}
    =
    \alpha(\hat{J}_x^2+\hat{J}_y^2)
    +
    p\hat{J}_z
    =
    \alpha(\Jtwo-\hat{J}_z^2)
    +
    p\hat{J}_z ,
    \label{eq:H_U1_form}
\end{equation}
so that
\([\hat{H},\hat{J}_z]=0 \).
Thus, the Hamiltonian conserves the total \(z\)-magnetization (charge) \(
    \hat{Q}
    =
    \sum_{i=1}^{N}
    \hat{\sigma}_z^{(i)}
    =
    2\hat{J}_z \).
Therefore, for \(\alpha_1=\alpha_2\), the same collective Hamiltonian conserves both the charge symmetry and the global permutation symmetry. We use Hamiltonian in Eq.~(\ref{eq:H_general}) to study permutation symmetry breaking dynamics, and Hamiltonian in Eq.~(\ref{eq:H_U1_form}) for charge symmetry breaking dynamics by selecting suitable symmetry-breaking initial states.

\subsection{Additive charge and non--additive permutation symmetry}
\label{subsec:additive_nonadditive}

The difference between the two symmetry structures, i.e., between charge symmetry and permutation symmetry, becomes clear when the system is bipartitioned into two subsystems \(A\) and \(B\). For the charge symmetry case, the conserved charge is additive. In other words,
\begin{equation}
    \hat{Q}
    =
    \hat{Q}_A\otimes \id_B
    +
    \id_A\otimes \hat{Q}_B, 
    \label{eq:Q_additive}
\end{equation}
where
\begin{equation}
    \hat{Q}_A
    =
    \sum_{i\in A}\hat{\sigma}_i^{z},
    \qquad
    \hat{Q}_B
    =
    \sum_{i\in B}\hat{\sigma}_i^{z} .
    \label{eq:QA_QB_def}
\end{equation}
Because of this additive structure, the total charge quantum number \(q\) is obtained by a simple additive relation. 
%Thus, the subsystem charge sectors combine directly to give the total charge sector.

However, the permutation symmetry in this sense is different. Although the collective spin vector is additive,
\begin{equation}
    \hat{\vec{J}}
    =
    \hat{\vec{J}}_A+\hat{\vec{J}}_B ,
    \label{eq:J_vector_additive}
\end{equation}
the square of the total spin 
\begin{equation}
    \Jtwo
    =
    \JAtwo \otimes \id_B
    +
    \id_A \otimes \JBtwo
    +
2\,\hat{\vec{J}}_A\cdot\hat{\vec{J}}_B \,,
    \label{eq:J2_nonadditive}
\end{equation}
is not additive. Here $(\hat{\vec{J}}_{A(B)} = (\,\hat J_{A(B),x} \,,\,\hat J_{A(B),y}\,
,\hat J_{A(B),z}\,)$ is the spin operator vector for subsystem \(A\,(\text{or}\,B)\). The cross term \(\hat{\vec{J}}_A\cdot\hat{\vec{J}}_B\) couples the two subsystems. Because of this term, the total spin quantum number
\(j\) is not obtained by a simple additive relation like in the charge case, but the subsystem spins \(j_A\) and \(j_B\) combine through angular-momentum
addition to give the allowed total-spin values \(j \in  \{|j_A-j_B|,\ (|j_A-j_B|+1),\ldots, (j_A+j_B)\} \). This distinction is central to our comparison of the two diagnostics for different symmetries as discussed later.

\subsection{Symmetry-breaking initial state}
\label{subsec:initial_states}
We now define the two initial states which are used to study the symmetry-breaking dynamics for the symmetries considered in this work. The first example breaks the additive charge symmetry, while the second example breaks the permutation symmetry in addition to the charge symmetry.

\subsubsection{Charge symmetry breaking state}

To probe charge symmetry-breaking dynamics, we use the tilted ferromagnetic product state, also known as the   \emph{spin coherent state}. It is given by \cite{Ares2023_EAoriginal,liHan2026_QME_U1,yamashika2026_QME_U1}
\begin{equation}
    |\Psi_0(\theta,\phi)\rangle
    =
    \bigotimes_{i=1}^{N}
    \left[
        \cos\frac{\theta}{2}|0\rangle_i
        +
        e^{i\phi}\sin\frac{\theta}{2}|1\rangle_i
    \right],
    \label{eq:charge_initial_state}
\end{equation}
with \(\theta\in [0,\pi]\) and \(\phi\in[0,2\pi]\) being the polar and azimuthal angles of the spins, respectively. For \(\theta=0\) or \(\pi\), the state \(\ket{\Psi_0}\) is an eigenstate of \(\hat Q\). We define reduced density state as $\rA = \Tr_B\left(\ket{\Psi_0}\bra{\Psi_0}\right)$. It can then be shown that  \([\rA , \QA] = 0\) for $\theta=0$ and $\pi$,  so that $\rA$ has a block-diagonal structure in subsystem charge basis of $\QA$, corresponding to possible \(q_A\) charge sectors. For other values of \(\theta\), it is a coherent superposition of different total charge sectors, and therefore breaks the charge symmetry generated by \(\hat{Q}\). Thus, \([\rA,\QA] \neq 0\) and \(\rA\) is no longer block-diagonal in \(\QA\) eigenbasis. In this work, we take \(\phi=0\) so that the tilt angle \(\theta\) controls the degree of initial charge symmetry breaking. Note that the initial state given in %Eq.~\eqref{eq:charge_initial_state} is also called \emph{spin coherent state} \cite{PhysRevE.97.052209}, Note that the state in 
Eq.~\eqref{eq:charge_initial_state} preserves the permutational symmetry. 

\subsubsection{Permutation symmetry breaking state}
To probe the permutational symmetry breaking dynamics, we prepare an inhomogeneous spin state by dividing the total spins into two equal parts and prepare them in different spin-coherent states \cite{Ieminietal} given by  
\begin{equation} 
    |\Psi_0(\theta_1,\theta_2)\rangle
    =
    \left[
        \bigotimes_{i=1}^{N/2}
        |\theta_1,0 \rangle_i
    \right]
    \otimes
    \left[
        \bigotimes_{i=N/2+1}^{N}
        |\theta_2,0 \rangle_i
    \right],
    \label{eq:perm_initial_state}
\end{equation}
where \( 
    |\theta,\phi\rangle
    =
    \left( \cos\frac{\theta}{2}|0\rangle
    +
    e^{i\phi}\sin\frac{\theta}{2}|1\rangle \right)\) is the bloch representation of a qubit \cite{nielsen2010quantum}, and we have set $\phi$ to zero for both the parts. In addition, we also set  $\theta_1=0, \, \theta_2=\Delta\theta$ so that $\Delta \theta$ measures the amount of permutation symmetry broken in the initial state. For example, when \(\Delta\theta=0\), all spins point in the same \((+z)\) direction, which lies entirely in the fully symmetric total-spin sector \(j=N/2\). In this fully symmetric sector, \([\rA,\JAtwo] = 0\) so that \(\rA\) is block-diagonal in \(\JAtwo\) eigenbasis. When \(\Delta\theta\neq0\), the two halves point in different directions. The state is then not invariant under permutations that exchange spins between the two halves. In other words, when expanded in the total-spin basis, the state has coherences between different spin sectors \(j\). This results to \([\rA,\JAtwo] \neq 0\) due to off-diagonal coherences between different subsystem spin sectors \(j_A\), provided subsystem $A$ has a mix of spins along $\theta_1$ and $\theta_2$. It is to be noted that, this initial state also breaks the charge symmetry.

\begin{comment}
Despite the breaking of global permutation invariance, permutation symmetry is retained within each subensemble. Writing the total collective spin as
\(
    \hat{\vec J}
    =
    \hat{\vec J}_1+\hat{\vec J}_2,
\)
the collective Hamiltonian Eq.~\eqref{eq:H_general} satisfies \(
    [\hat H,\hat{ J}_1^2]
    =
    [\hat H,\hat{J}_2^2]
    = 0.\)
Therefore, quantum numbers $j_1$ and $j_2$ associated
with the two subensembles are constants of motion. Since each subensemble
is initially prepared in its fully symmetric sector,
\(j_1={N_1}/2\,,\,j_2=N_2/2\),
the dynamics remains restricted to the product Hilbert space \(
    \mathcal H_{j_1}\otimes\mathcal H_{j_2}
\) with dimension \((N_1+1)(N_2+1)\) \cite{Ieminietal},
and
inhomogeneous initial condition with $\Delta\theta\neq0$ can contain
coherences between several global $j$ sectors.
\end{comment}

\begin{figure*}[t]
    \centering

    % ---------------- (a) Top figure ----------------
    \includegraphics[
        width=0.83\textwidth
    ]{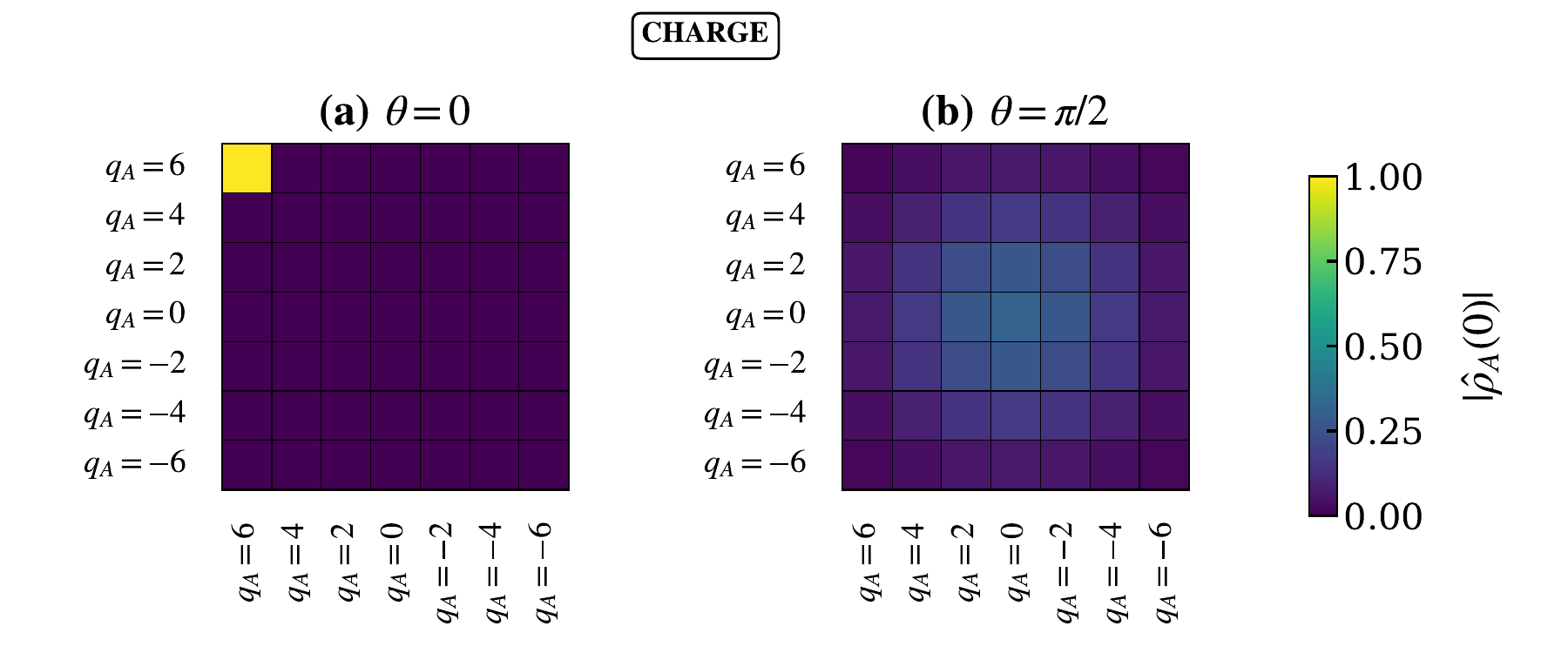}
    \vspace{0.35cm}

    % ---------------- (b) Bottom figure ----------------
    \includegraphics[
        width=0.83\textwidth
    ]{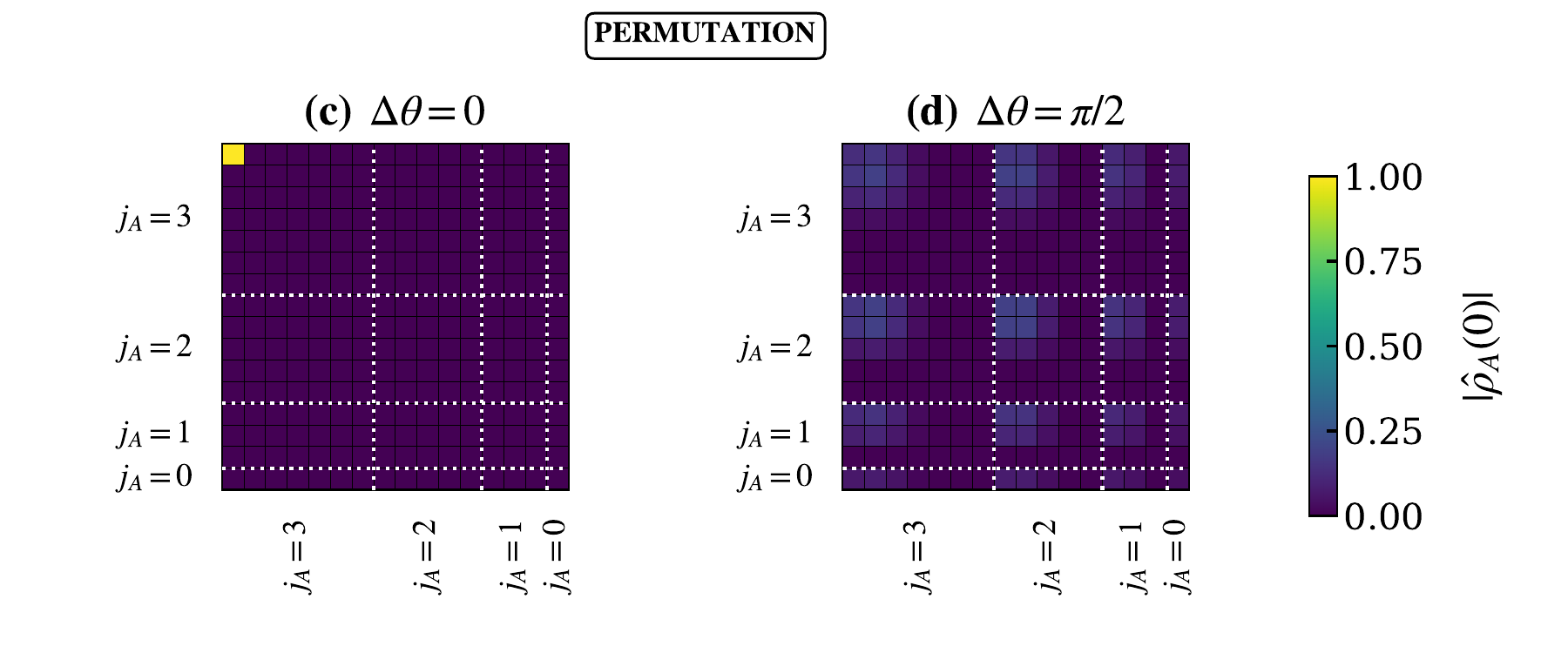}

    \caption{
Structure of the initial reduced density matrix $\rA(0)$ in the symmetry-resolved basis for the charge and permutational-symmetry cases.
Panels \textbf{(a,b)} show magnitude of $\rA(0)$, denoted as $|\rA(0)|$, in the subsystem-charge $q_A$ basis for the tilted ferromagnetic initial state, as given in Eq.~\eqref{eq:charge_initial_state} with (a) $\theta=0$, and (b) $\theta=\pi/2$, respectively, demonstrating support over several $q_A$ sectors when $\theta=\pi/2$ .
%For $\theta=0$, the state occupies a single charge sector, whereas for $\theta=\pi/2$ it has support over several $Q_A$ sectors.
Panels \textbf{(c,d)} show the corresponding reduced density matrix in the subsystem total-spin $j_A$ basis for the initial state  given in Eq.~\eqref{eq:perm_initial_state} with (c) $\Delta\theta=0$ and (d) $\Delta\theta=\pi/2$, respectively, with $\Delta \theta=\pi/2$ showing support
and coherences across different $j_A$ sectors. 
%See main text for more details.
%For $\Delta\theta=0$, the state is confined to the fully symmetric $j_A$ sector, while for $\Delta\theta\neq0$ it develops support and coherences across different $j_A$ sectors, reflecting the breaking of permutation invariance in the initial state. 
These plots correspond to \(N = 80\,,N_A = 6\), where subsystem A consists of spins from site $38$ to $43$.
}
    \label{fig:rhoA0_contours}
\end{figure*}

Fig.~(\ref{fig:rhoA0_contours}) shows how the structure of the initial reduced density matrix changes as the corresponding symmetry is broken. For the charge-symmetry case, Fig.~(\ref{fig:rhoA0_contours}a) shows that at $\theta=0$ the subsystem state is confined to a single $q_A$ sector, consistent with an initial state that preserves the charge symmetry. In contrast, for $\theta=\pi/2$ [see Fig.~(\ref{fig:rhoA0_contours}b)], the reduced density matrix acquires population over several $q_A$ sectors together with off-diagonal coherences between them, indicating charge--symmetry breaking. A similar behavior is observed for the permutational-symmetry case. For $\Delta \theta = 0$ [Fig.~(\ref{fig:rhoA0_contours}c)], all spins are aligned in $(+z)$ direction and the reduced state is restricted to the fully symmetric $j_A = N_A/2$ sector. When  $\Delta \theta \ne 0$, say $\pi/2$ [Fig.~(\ref{fig:rhoA0_contours}d)], the state develops coherences across different $j_A$ sectors. Thus, Fig.~(\ref{fig:rhoA0_contours}) directly illustrates how varying $\theta$ and $\Delta \theta$ introduces symmetry-breaking coherences, thereby serving as a symmetry-breaking parameter for  charge and permutation symmetries, respectively.

%===============================

% CONTOUR PLOTS FOR CHARGE AND PERMUTATION SYMMETRY

%===============================

\subsection{Symmetry-based and energy-based diagnostics}
\label{subsec:diagnostics}
Once the system is prepared in the desired initial state, it will evolve as
\( |\Psi(t)\rangle = e^{-i\hat{H}t}|\Psi_0\rangle \). 
We now define the two diagnostics used throughout the paper to study the symmetry breaking dynamics. The definitions are given in a way that is independent of the symmetry under consideration.\\

\subsubsection{Entanglement asymmetry: symmetry based diagnostic}
This diagnostic is predominantly used to probe \rm{QME} \cite{Ares2023_EAoriginal,Ares2023_lack,Ferro_2024}, where a specific symmetry of the Hamiltonian is broken by an initial state. Let \(\hat{\mathcal{C}}_A\) be the symmetry operator corresponding to the subsystem \(A\). For example, to probe the charge symmetry,
\(\hat{\mathcal{C}}_A=\hat{Q}_A \), and to probe the permutational symmetry, \(\hat{\mathcal{C}}_A=\JAtwo\).
Let \(\hat{\mathcal{P}}_{\lambda}^{A}\) be the projector onto the eigenspace of \(\hat{\mathcal{C}}_A\) with eigenvalue \(\lambda\). Then,
\begin{equation}
   \rACA \,=\,
    \sum_{\lambda}
    \hat{\mathcal{P}}_{\lambda}^{A} \,
    \hat{\rho}_A \,
    \hat{\mathcal{P}}_{\lambda}^{A} \,,
    \label{eq:rho_dephased_general_def}
\end{equation}
is the \emph{symmetry-dephased reduced state} corresponding to the subsystem symmetry operator.
For charge symmetry, \(\lambda\) labels the subsystem charge sectors \(q_A\) and for the total-spin case, \(\lambda\) labels the possible subsystem spin sectors \(j_A\).

The entanglement asymmetry is then defined as 
\cite{Ares2023_EAoriginal}
\begin{equation}
    \Delta S_A
    =
    S\left(
        \rACA
    \right)
    -
    S\left(
        \hat{\rho}_A
    \right),
    \label{eq:EA_def}
\end{equation}
where \(S(\hat{\rho}) = -\Tr(\hat{\rho}\ln\hat{\rho})\) is the \emph{von-Neumann entropy} of the state $\hat \rho$. This diagnostic is purely symmetry based as it measures the entropy increase caused by removing coherences between the chosen subsystem symmetry sectors. In particular, \(\Delta S_A(t)=0\) if and only if \(\hat{\rho}_A(t)\) is block diagonal in the eigenspaces of \(\hat{\mathcal{C}}_A\) \textit{i.e,} when \(\rA =\rACA\).  \\

%\textcolor{blue}{can remove relative entropy discussion}Also, \(\EA\) is equal to the \emph{relative entropy measure} between \(\rA(t)\) and its symmetry-dephased counterpart \cite{Ares_2023}

%\begin{eqnarray}
%    \EA
%    &=&
%    S\left(\rA\,\middle\|\,\rACA\right) \nonumber \\
%    &=& \Tr\!\left[\rA
%\left(
%\ln \rA
%-
%\ln \rACA
%\right)
%\right],
    %\label{eq:EA_relative_entropy}
%\end{eqnarray}

%which is always positive ($\EA \geq 0$) \cite{nielsen2010quantum}. 

\subsubsection{Trace distance: energy-based diagnostic}
Several other literatures have considered certain energy-based diagnostics as a probe for Mpemba-like relaxation in quantum systems, which does not consider any symmetry present in the system \cite{noGlobal,Mpemba_twolevelsystem,zhou2026_quasi}. One such measure is the trace distance which we define below. 

Let \(\{|E_n\rangle\}\) be the eigenbasis of a nondegenrate Hamiltonian \(\hat{H}\). Then, the reduced diagonal ensemble \(\rADE\) is defined as \cite{noGlobal,diag_ensemble_2021_cirac} 
\begin{eqnarray}
\rADE &=& \Tr_B \left(\rDE\right) \nonumber\\
    %&=&\sum_n \langle E_n |\, \hat \rho(t=0)\, | E_n \rangle \,\Tr_B \!\left( |E_n\rangle\langle E_n| \right) \nonumber \\ 
      &=&  \sum_n |c_n|^2  \,\Tr_B \!\left( |E_n\rangle\langle E_n| \right) \,,
    \label{eq:rhoADE_def}
\end{eqnarray}
where \(\rDE = \sum_n |c_n|^2 \,\!\left( |E_n\rangle\langle E_n| \right)\) is the diagonal ensemble for the full system,  and \(c_n=\langle E_n|\Psi_0\rangle\). In case of a degenerate Hamiltonian, the projection operator onto each eigenvalue should take into account all the degerate eigenstates and each block in the block-form of $\rDE$ corresponds to different energy eigen subspaces \cite{diag_ensemble_2021_cirac}. The trace distance \(\Dtd(t)\) is then defined as \cite{nielsen2010quantum}
\begin{equation}
    \Dtd(t)=\frac{1}{2}\left\|\hat{\rho}_A(t) - \hat{\rho}_{\rm DE}^{A}\right\|_1 \, , 
\label{eq:TD_def}
\end{equation}
where \(\|\hat A\|_1 = \Tr\left(\sqrt{ \hat{A}^{\dagger}\hat{A}} \right)\) is the trace norm. This diagnostic is purely energy based as it compares the instantaneous reduced state with the reduced state obtained after removing energy-basis coherences in the full system. It does not require specifying a symmetry. Therefore, \(\Dtd(t)\) probes relaxation toward the reduced diagonal ensemble rather than symmetry restoration directly.

\begin{figure*}[h]
    \centering

    \begin{tabular}{cc}
        \includegraphics[width=0.40\textwidth]{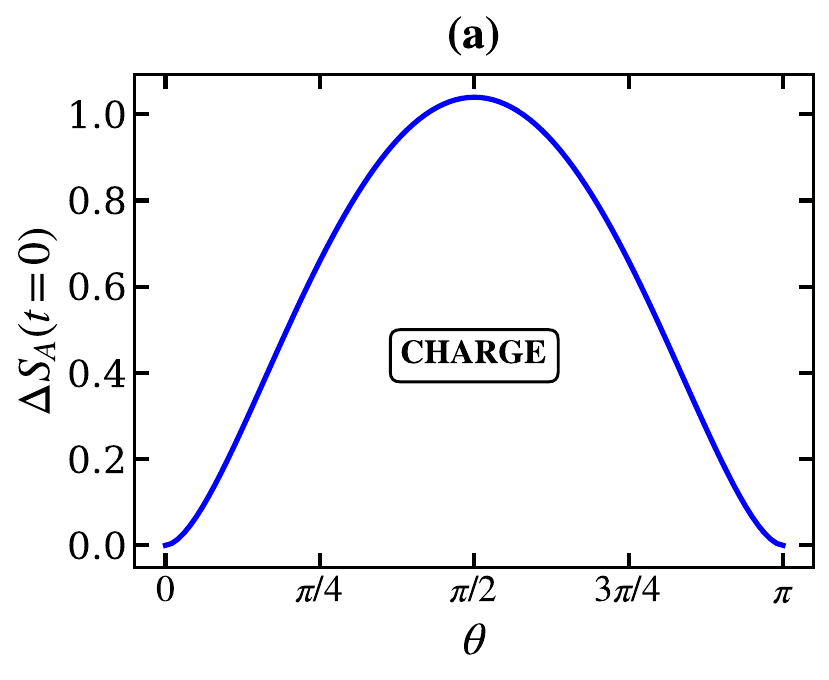} 
        &
        \includegraphics[width=0.41\textwidth]{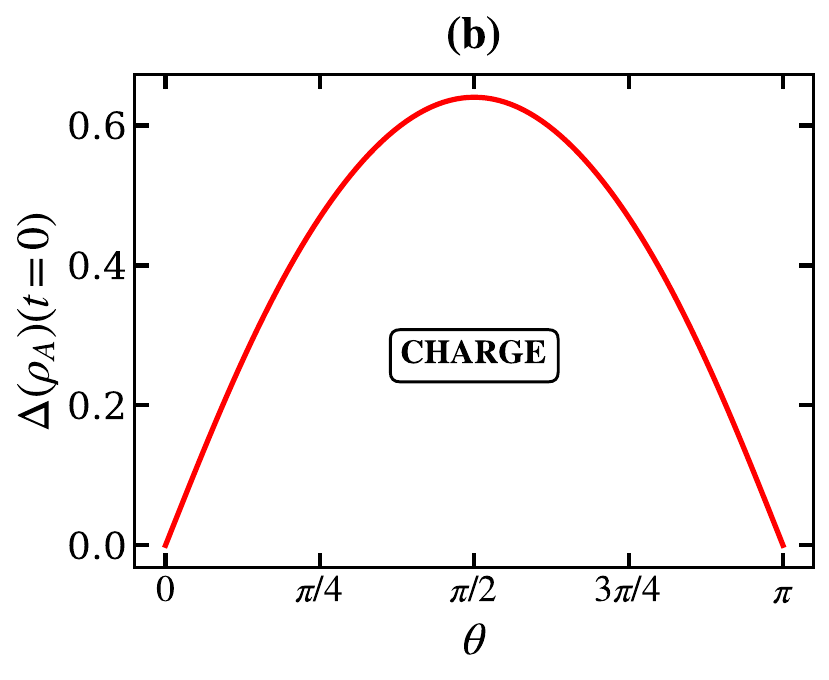}
        \\
        \\[3mm]
        \includegraphics[width=0.40\textwidth]{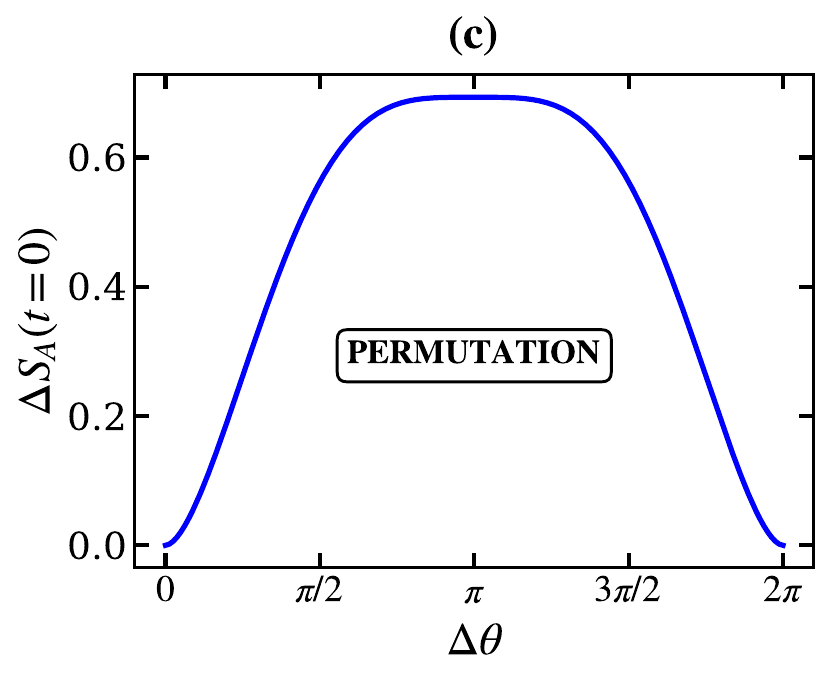}
        &
        \includegraphics[width=0.40\textwidth]{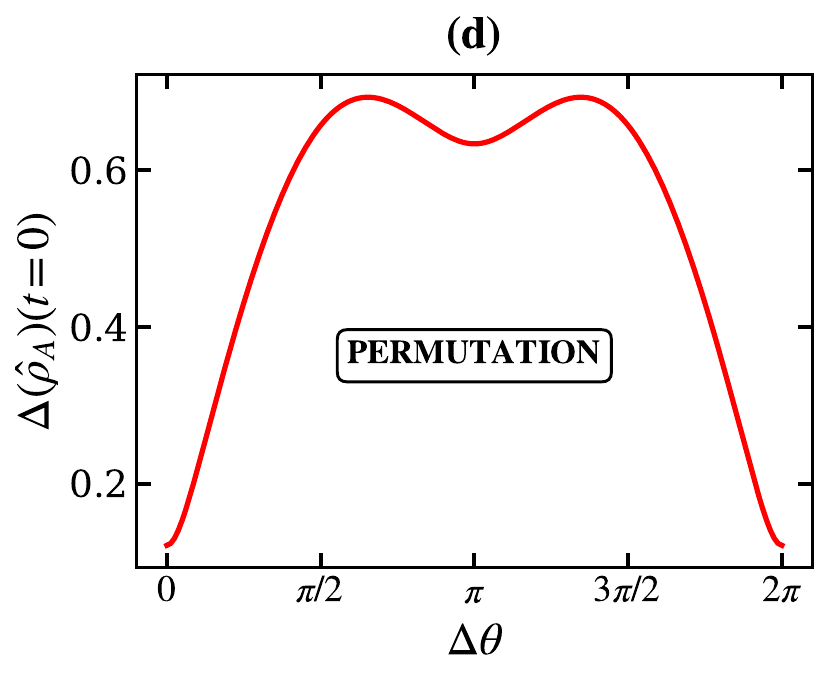}
        \\
    \end{tabular}

    %\caption{
    %Initial values of the entanglement asymmetry and trace distance for
    %charge and permutational symmetry-breaking initial states : 
    %(a) Initial entanglement asymmetry \(\Delta S_A(t=0)\) and (b) Initial trace distance \(\Delta(\rho_A)(t=0)\) as a function
    %of \(\theta\) for the charge-symmetry breaking initial state.
    %(c) Initial entanglement asymmetry \(\Delta S_A(t=0)\) and (d) Initial trace distance \(\Delta(\rho_A)(t=0)\) as a function
    %of the polar angle difference \(\Delta\theta\) for the permutational-symmetry breaking initial state
    %}
    \caption{Initial values of the entanglement asymmetry \(\EA(t\!=\!0)\) and trace distance \(\Dtd(t\!=\!0)\) for
    charge and permutation-symmetry-breaking initial states. 
    Panels \textbf{(a,b)} correspond to the charge symmetry-breaking initial state (Eq.~\eqref{eq:charge_initial_state}), and the diagnostics are
    plotted as a function of the tilt angle \(\theta\). The reduced diagonal ensemble
    \(\rho_{\mathrm{DE}}^A\) is constructed with respect to the
    charge-symmetric Hamiltonian in Eq.~\eqref{eq:H_U1_form} with
    \(\alpha=1/N\) and \(p=\pi/2\).
    Panels \textbf{(c,d)} correspond to the permutational symmetry-breaking initial state in  Eq.~\eqref{eq:perm_initial_state}
    where the diagnostics are plotted as functions of the polar angle
    difference \(\Delta\theta\). Here,
    \(\rho_{\mathrm{DE}}^A\) is constructed with respect to Hamiltonian in Eq.~\eqref{eq:H_general}, with
    \(\alpha_1=1/N\), \(\alpha_2=2/N\), and \(p=\pi/2\). In both figures, \(N=80\) and \(N_A=2\) with subsystem $A$ consisting of spins at site $40$ and $41$. .
    }
\label{fig:initial_EA_TD_comparison}
\end{figure*}

\subsection{Diagnostic comparison and objective}
\label{subsec:diagnostic_objective}
The two diagnostics introduced above answer different questions. The trace distance \(\Dtd(t)\) asks how close the subsystem is to the reduced diagonal ensemble \(\rADE\), and is therefore tied to relaxation in the energy eigenbasis. The entanglement asymmetry \(\Delta S_A(t)\) asks how much coherence remains between different symmetry sectors, and is therefore only tied to symmetry restoration dynamics.

The central aim of this work is to determine whether these two diagnostics give the same temporal information about quantum Mpemba-like relaxation. We compare the behavior in two cases:
\begin{enumerate}[label=(\roman*)]
    \item a charge symmetry-breaking initial state evolved under a Hamiltonian with charge conservation;
    \item a permutation symmetry-breaking initial state evolved under a permutation symmetric Hamiltonian with global \(\Jtwo\) conservation.
\end{enumerate}
Below, we briefly discuss the criteria for QME: 
Let two initial states, \(\hat\rho_0\) and \(\hat\rho'_0\), evolve under
the same Hamiltonian dynamics as
\[
    \hat\rho(t)=e^{-i\hat Ht} \, \hat  \rho_0 e^{i\hat Ht},
    \qquad
    \hat\rho'(t)=e^{-i\hat Ht}\,\hat\rho'_0 e^{i\hat Ht}.
\]
Let \(\mathcal{O}[\hat\rho(t)]\) denote a diagnostic that measures the
distance of the time-evolved state from the corresponding stationary or
symmetry-restored reference state such that initially
    $\mathcal{O}[\hat\rho'_0]
    <
    \mathcal{O}[\hat\rho_0]$.
A \rm{QME} is said to occur with respect
to the diagnostic \(\mathcal{O}\) if, after some finite time \(\tau_m\), the
ordering reverses, i.e., $\mathcal{O}[\hat\rho'(t)] > \mathcal{O}[\hat\rho(t)]$ for $t>\tau_m $,
where \(\tau_m\) can be referred as \emph{crossover or Mpemba time} for those states \cite{parityTime_QME}. This means that the state which was initially farther from the reference
state relaxes faster and becomes closer to the reference state at later times.

% 1) CHARGE SYMMETRY 

\begin{figure*}[h]
\centering

\begin{minipage}[h]{0.495\textwidth}
    \centering
    \includegraphics[width=\linewidth]{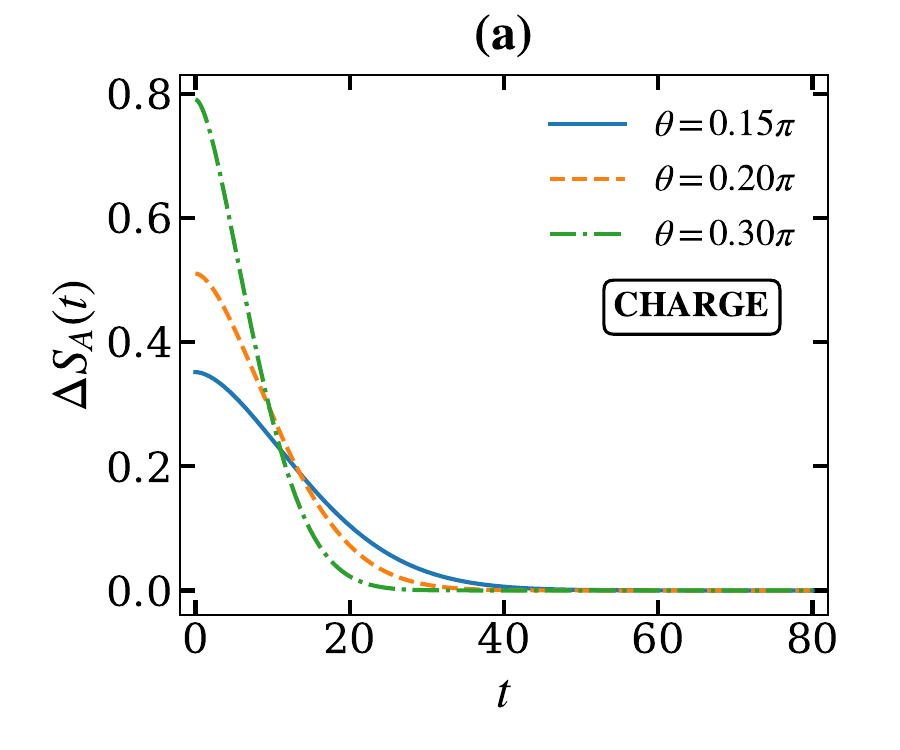}
    \vspace{1mm} 

\end{minipage}
\hfill
\begin{minipage}[h]{0.495\textwidth}
    \centering
    \includegraphics[width=\linewidth]{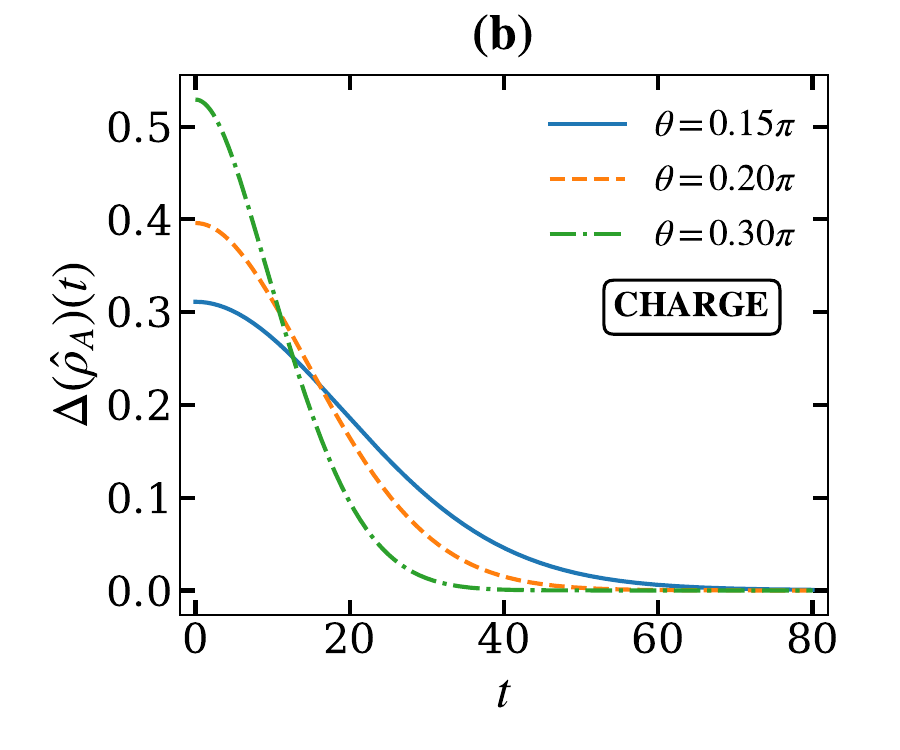}
    \vspace{1mm}

\end{minipage}

\caption{Charge symmetry-breaking dynamics : 
\textbf{(a)} Time evolution of the  entanglement asymmetry
\(\Delta S_A(t)\) and the \textbf{(b)} trace distance
\(\Delta(\rho_A)(t)\) for charge symmetry case.
The three curves correspond to different symmetry-breaking parameters, $i.e.$ initial tilt angle \(\theta\). 
For larger \(\theta\), the initial value of both
diagnostics is larger but the decay is faster. The results are obtained by exact diagonalization for \(N=80\,,\,N_A = 2\), with the Hamiltonian in Eq.~\eqref{eq:H_U1_form} and parameters set to \(\alpha = 1/N\,,\,p = \pi/2\)
}
\label{fig:charge_TD_EA}

\end{figure*}

\begin{comment}

\textcolor{blue}{
\subsection{Criteria of Quantum Mpemba effect}
\label{subsec:criteria_QME}
The quantum Mpemba effect can be formulated as an anomalous reversal in
the relaxation ordering of two initially nonequilibrium quantum states, as discussed below.\\
Let two initial states, \(\hat\rho_0\) and \(\hat\rho'_0\), evolve under
the same Hamiltonian dynamics as
\[
    \hat\rho(t)=e^{-i\hat Ht}\hat\rho_0 e^{i\hat Ht},
    \qquad
    \hat\rho'(t)=e^{-i\hat Ht}\hat\rho'_0 e^{i\hat Ht}.
\]
Let \(\mathcal{D}[\hat\rho(t)]\) denote a diagnostic that measures the
distance of the time-evolved state from the corresponding stationary or
symmetry-restored reference state. If the initial ordering is such that
\begin{equation}
    \mathcal{D}[\hat\rho'_0]
    <
    \mathcal{D}[\hat\rho_0],
    \label{eq:QME_initial_ordering}
\end{equation}
then \(\hat\rho_0\) is initially farther from the reference state than
\(\hat\rho'_0\). A quantum Mpemba effect is said to occur with respect
to the diagnostic \(\mathcal{D}\) if, at some finite time \(\tau_m\), the
ordering reverses,
\begin{equation}
    \mathcal{D}[\hat\rho'(t)]
    >
    \mathcal{D}[\hat\rho(t)],
    \qquad
    t>\tau_m .
    \label{eq:QME_ordering_reversal}
\end{equation}
where \(\tau_m\) can be referred as \emph{crossover or Mpemba time} for those states. This means that the state which was initially farther from the reference
state relaxes faster and becomes closer to the reference state at later times.}

\end{comment}

Fig.~(\ref{fig:initial_EA_TD_comparison}) shows the initial ordering of
the states according to the two diagnostics. For charge symmetry, the
symmetry-breaking parameter is the tilt angle \(\theta\), while for
permutational symmetry it is the polar-angle difference
\(\Delta\theta\) as discussed in Sec \ref{subsec:initial_states}. In both cases, increasing the corresponding parameter
increases the initial value of \(\EA(0)\) and \(\Dtd(0)\) upto some parameter value,  and is symmetric about \(\theta=\pi/2\)  for charge symmetry, and \(\Delta\theta=\pi\) for permutation symmetry in the parameter range studied. These plots
therefore establish which initial states are farther from the
symmetry-restored or stationary reference state. The \rm{QME} is then detected from the subsequent time evolution when the state with
the initially larger diagnostic value decays faster and crosses the state with the initially smaller
one.

%and symmetric about value of \(\theta\) or \(\Delta\theta\) of maximal considered symmetry breaking

% ============================================================
\section{Numerical Results}
\label{sec:numerical_observation}
% ============================================= 
In this section, we numerically compare the time evolution of the entanglement asymmetry \(\EA(t)\) and trace distance \(\Dtd(t)\) for both charge and permutation symmetry breaking initial states, evolved with appropriate symmetric Hamiltonians, and probe the QME in these systems. For numerical calculations in the rest of the paper, we have set $N=80$, $N_A=2$ with subsystem $A$ consisting of spins at site $40$ and $41$. In order to simulate larger system sizes, we exploit the collective nature of the Hamiltonian as discussed below. For a  general $N$ spin system with $N_1$ spins along one direction and $N_2$ spins along some other direction, the total collective spin operator 
$\J$ can be written as
\(
    \hat{\vec J}
    =
    \hat{\vec J}_1+\hat{\vec J}_2.
\)
 It can be shown that the Hamiltonian in Eq.~\eqref{eq:H_general} commutes with $\hat J_1^2$ and $\hat J_2^2$ separately, i.e., 
 \(
    [\hat H,\hat{ J}_1^2]
    =
    [\hat H,\hat{J}_2^2]
    = 0.\)
Therefore, the dynamics remain restricted to the product Hilbert space of dimension \((N_1+1)(N_2+1)\) \cite{Ieminietal}, enabling the study of comparatively larger system sizes. Note that for the state given in Eq.~\eqref{eq:perm_initial_state}, $N_1=N_2=N/2$.
% ============================================================
\subsection{Charge symmetry}
\label{subsec:numerical_charge_symmetry}
% ============================================================

Fig.~(\ref{fig:charge_TD_EA}a) shows the time evolution of the entanglement asymmetry \(\EA(t)\) for different tilt angles \(\theta\) in the case of charge symmetry-breaking initial state (\textit{tilted ferromagnetic state}) given in Eq. \eqref{eq:charge_initial_state}, which is evolved under charge-symmetric Hamiltonian in Eq.~\eqref{eq:H_U1_form}. The initial value \(\EA(t\!=\!0)\) increases with \(\theta\) [see Fig.~(\ref{fig:initial_EA_TD_comparison})], which
indicates larger symmetry breaking in larger tilt angles. On the other hand, as seen from Fig.~(\ref{fig:charge_TD_EA}a), its decay with time is 
faster for larger \(\theta\). This is
the characteristic signature of Mpemba-like relaxation, where the
state with stronger initial charge-symmetry breaking restores the
subsystem charge symmetry faster. We also see that \(\EA(t) \rightarrow 0\) for larger times, which indicates that the symmetry in the subsystem is being restored. Fig.~(\ref{fig:charge_TD_EA}b) shows the corresponding time evolution
of the trace distance \(\Dtd(t)\) for the same tilt angles \(\theta\). 
We observe that their relaxation behaviour also have the same qualitative
temporal behaviour, and thus could also clearly probe the symmetry--based quantum Mpemba effect in this system. 

The crucial observation is that both \(\EA(t)\) and \(\Dtd(t)\) show the same temporal and crossing behavior.
We argue later that, in the charge symmetry case, the trace distance
dynamics is dominated by the decay of the off-diagonal charge sector
coherence. Since the entanglement asymmetry directly measures such
coherence, both diagnostics capture the same dynamical symmetry restoration.

% ============================================================
\subsection{Permutational symmetry}
\label{subsec:numerical_permutation_symmetry}
% ============================================================

\begin{figure*}[t]
\centering

\begin{minipage}[t]{0.495\textwidth}
    \centering
    \includegraphics[width=\linewidth]{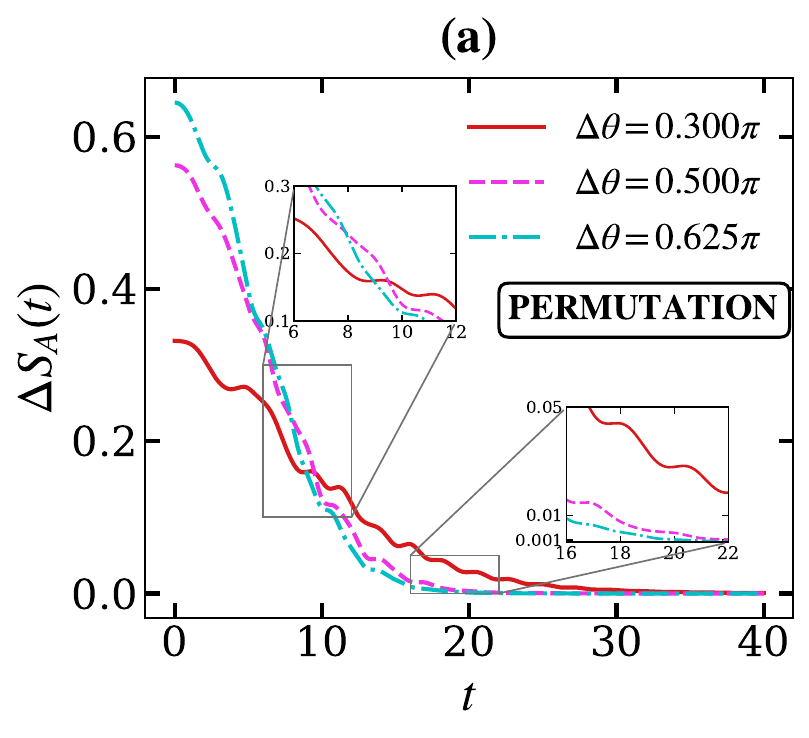}
    \vspace{1mm}

\end{minipage}
\hfill
\begin{minipage}[t]{0.495\textwidth}
    \centering
    \includegraphics[width=\linewidth]{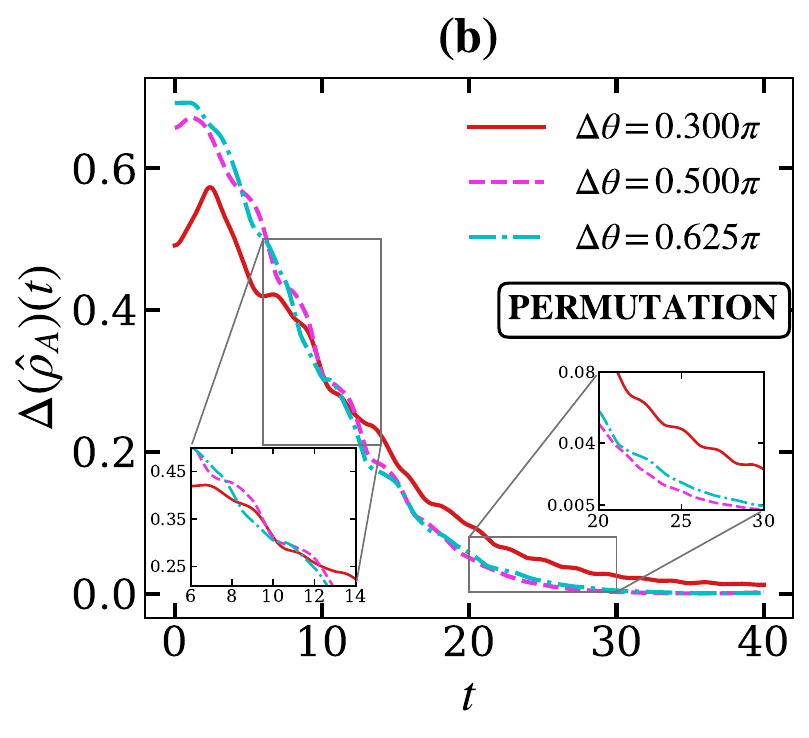}
    \vspace{1mm}
\end{minipage}

\caption{Permutational symmetry-breaking dynamics : 
\textbf{(a)} Time evolution of the entanglement asymmetry
\(\Delta S_A(t)\) and \textbf{(b)} trace distance
\(\Delta(\rho_A)(t)\) for the permutational symmetry case.
The three curves correspond to different symmetry-breaking parameter \(\Delta\theta\). 
The results are obtained by exact diagonalization for  \(N=80\,,\,N_A=2\), by drawing subsystem from the middle of the spin chain,
for the Hamiltonian in Eq.~\eqref{eq:H_general} with parameters
\(\alpha_1=1/N\), \(\alpha_2=2/N\), and \(p=\pi/2\).
The insets magnify the indicated intermediate and late time regions to clearly highlight the differences between the curves.
}
\label{fig:perm_TD_EA}

\end{figure*}

Fig.~(\ref{fig:perm_TD_EA}a) shows the time evolution of the
entanglement asymmetry \(\EA(t)\), computed with respect to the
subsystem total-spin operator \(\JAtwo\), for different initial polar
angle differences \(\Delta\theta\). Here the initial state given in Eq.~\eqref{eq:perm_initial_state} breaks the permutational symmetry, and is evolved under the permutationally symmetric Hamiltonian in Eq.~\eqref{eq:H_general}. The corresponding asymmetry is measured by
removing the coherences between different \(j_A\) sectors of the reduced
density matrix. Similar to the charge symmetry breaking dynamics discussed above,
\(\EA(t)\) also shows a clear Mpemba--like relaxation in the
permutation symmetry case. The initial value of \(\EA(t)\) increases
with  \(\Delta \theta\), indicating that larger \(\Delta \theta\)
corresponds to stronger initial coherence between different subsystem
\(j_A\) sectors. As seen from Fig.~(\ref{fig:perm_TD_EA}a), the higher asymmetric initial states relax
faster, thus demonstrating quantum Mpemba relaxation. 

Figure~(\ref{fig:perm_TD_EA}b) shows the corresponding time evolution
of the trace distance \(\Dtd(t)\). Although \(\Dtd(t)\) also decreases
with times, its behavior is not qualitatively similar to that of \(\EA(t)\). We observe multiple crossings between the curves corresponding to different $\Delta \theta$ during the initial times. In addition, as shown in the magnified inset of Fig. (\ref{fig:perm_TD_EA}b), higher initial $\Delta \theta$ (blue line) continues to have larger values of trace distance as compared to smaller $\Delta \theta$ at larger times, contradicting the Mpemba like behavior. Therefore, unlike the charge-symmetry case, trace distance \(\Dtd(t)\) does not provide a clean probe of the 
Mpemba-like relaxation behavior as observed in \(\EA(t)\).
Hence, in the permutational symmetry-case, the entanglement asymmetry remains a better diagnostic of Mpemba effect. The trace distance,
however, is a broader relaxation diagnostic and includes additional
energy-based relaxation effects.

%==========================================================

% MAIN PLOTS FOR QME IN CHARGE AND PERMUTAIION SYMMETRY

%---------------%
% 2) PERMUTATIONAL SYMMETRY

%=========================================================

\section{Analytical Discussion}
\label{sec:analytical}
In this section, we provide analytical arguments behind the different dynamical behaviors observed numerically in the two diagnostics for the case of charge and permutation symmetry.
To find the relation between \(\EA(t)\) and \(\Dtd(t)\), we can write the difference between \(\rA\) and \(\rADE\) as follows :
\begin{equation}
\!\!\rA(t)-\rADE
    \!=\!
    \left[
        \rA(t)- \hat \rho_{A,\hat{\mathcal{C}}_A}(t)
    \right]
    \!+\!
    \left[
         \hat \rho_{A,\hat{\mathcal{C}}_A}(t) - \rADE
    \right].
    \label{eq:W_delta_decomposition}
\end{equation}
Let us define the operator
\begin{equation}
    \Wop(t)=\rA(t)- \hat \rho_{A,\hat{\mathcal{C}}_A}(t)
    \label{eq:W_def}
\end{equation}
which contains the off-block-diagonal coherences between subsystem symmetry sectors. It is precisely the symmetry dephased part removed from the reduced state. Therefore, \(\Wop(t)\) is the operator that directly encodes subsystem symmetry breaking. The second term,
\begin{equation}
    \dop(t)= \hat \rho_{A,\hat{\mathcal{C}}_A}(t)-\rADE \, ,
    \label{eq:delta_def}
\end{equation}
is the residual difference between the symmetry-dephased reduced state and the reduced diagonal ensemble. 
Similar kind of decomposition has been previously discussed from a resource--theoretic perspective in Ref.~\cite{unification2026_decomposition}. 
Using Eq.~\eqref{eq:W_delta_decomposition}, the trace distance becomes
\begin{equation}
    \Dtd(t)
    =
    \frac{1}{2}\normone{\Wop(t)+\dop(t)} \,.
    \label{eq:TD_W_delta}
\end{equation}
Employing the triangle inequality, we obtain \begin{equation}
    \Dtd(t)
    \leq
    \frac{1}{2}\normone{\Wop(t)}
    +
    \frac{1}{2}\normone{\dop(t)},
    \label{eq:TD_upper_bound}
\end{equation}
while the reverse triangle inequality gives
\begin{equation}
    \Dtd(t)
    \geq
    \frac{1}{2}
    \left| \;
        \normone{\Wop(t)}
        -
        \normone{\dop(t)}
    \;\right| \,.
    \label{eq:TD_lower_bound}
\end{equation}
Therefore, \(\Dtd(t)\) is bounded by sum and differences of individual trace norms of \(\Wop(t)\) and \(\dop(t)\). It will then track the symmetry-breaking dynamics present in \(\Wop(t)\) only if the time dependence of \(\dop(t)\) is zero or very weak compared to that of \(\Wop(t)\). 
In the case of \(\EA(t)\), it measures the entropy increase of removing the off-sector symmetry coherence from \(\rA(t)\), thus directly connected to only \(\Wop(t)\).
Thus, we conclude that \(\EA(t)\) and \(\Dtd(t)\) will show similar relaxation behavior when \(\dop(t)\) vanishes or is weakly time dependent.
%The above discussion highlights the conditions for similar relaxation behavior of entanglement asymmetry and trace distance.

%=================================================

%delta_A(t) = 0 justification 

\subsection{Exact vanishing of \(\dop(t)\) in the collective charge symmetry model}
\label{subsec:delta_exact_collective_charge}
%\sout{We now consider the charge conserving limit for our collective Hamiltonian as given in Eq.~\eqref{eq:H_U1_form}.}
In this subsection, we show that for the spin-coherent initial states given in Eq.~\eqref{eq:charge_initial_state}, the residual contribution \(\dop(t)\) vanishes at all times, for the collective
charge-symmetric Hamiltonian in  Eq.~\eqref{eq:H_U1_form}. Consequently, the trace distance \(\Dtd(t)\) is determined entirely by the charge-coherence contribution \(\Wop(t)\). 

Since \([\hat{H},\Q]=0\), the diagonal ensemble $\rDE$ commutes with the total charge
\( [\rDE,\Q]=0 \).
Taking the partial trace over the subsystem \(B\), and using the additive form in Eq.~\eqref{eq:Q_additive}, we obtain 
\begin{equation}
    [\rADE,\QA]=0 \,.
    \label{eq:rhoADE_commutes_QA}
\end{equation}
Thus, the reduced diagonal ensemble \(\rADE\) and therefore, \(\dop(t) = (\rQA(t) - \rADE)\) is block diagonal in the subsystem charge basis \(\QA\). 
%We now show that this residual term vanishes exactly for the collective charge-symmetric Hamiltonian Eq.~\eqref{eq:H_U1_form} with a spin coherent initial state Eq.~\eqref{eq:charge_initial_state}, which breaks the charge symmetry.

As shown in section \ref{subsec:collective_hamiltonian}, the Hamiltonian in Eq.~\eqref{eq:H_U1_form}
is diagonal in the simultaneous eigenbasis of \(\Jtwo\) and
\(\hat J_z\). Since the spin coherent initial state is fully permutation
symmetric, it has support only in the maximum-spin sector \(j=N/2\), and can be
expanded in the Dicke basis \({\ket{D_k^N}}\). The Dicke basis is defined as \cite{Seshadri2018}
\begin{equation}
    \ket{D_k^N}
    =
    \frac{1}{\sqrt{\binom{N}{k}}}
    \sum_{\mu:\,w(\mu)=k}
    \ket{\mu},
    \label{eq:Dicke_define}
\end{equation}
where \(w(\mu)=k\) denotes the Hamming weight of the computational basis
state \(\ket{\mu}\). In the present convention, \(k\) is the number of down
spins (\(-z\) direction) in a \(N\)-spin system \cite{Dicke1954,Seshadri2018,Ieminietal}, and \(\binom{N}{k} = {N!}/{k!(N-k)!}\) is the binomial coefficient. Therefore,
\begin{equation}
\ket{D_k^N}\equiv\ket{j,m_k},
\quad \mathrm{with}~~
j=N/2,
\quad
\mathrm{and}~~m_k=j-k .
\label{eq:Dicke_jm_intro}
\end{equation}
The total charge eigenvalue is
$q_k=N-2k=2 m_k$.
%\label{eq:charge_eigenvalue_dicke}
%\end{equation}
Using the angular-momentum eigenvalue relations, one obtains
\begin{equation}
\hat H\ket{D_k^N}
\!=\!
E_k\ket{D_k^N},
\,\,
E_k
=
\alpha\left[j(j+1)\!-\!m_k^2\right]\!+\!p\,m_k .
\label{eq:dicke_energy}
\end{equation}
The spin coherent state in  Eq.~\eqref{eq:charge_initial_state} can
be expressed in terms of the Dicke states as 
\begin{equation}
\ket{\Psi_0(\theta, \phi)}
=
\sum_{k=0}^{N}
A_k (\theta, \phi) \, \ket{D_k^N},
\label{eq:SCS_dicke_expansion}
\end{equation}
with
\begin{equation} \label{eq:SCS_dicke_coeff}
A_k(\theta, \phi)
=
\sqrt{\binom{N}{k}}
\left(\cos\frac{\theta}{2}\right)^{N-k}
\left(e^{i\phi}\sin\frac{\theta}{2}\right)^k.
\end{equation}
The time evolution of the state is then given by
\begin{equation}
\ket{\Psi(t)}
=
\sum_{k=0}^{N}
A_k (\theta, \phi) \, e^{-iE_k t} \, \ket{D_k^N}.
\label{eq:SCS_dicke_time_evolve}
\end{equation}
Thus, the time evolution preserves the Dicke-state populations and only introduces relative phases between different Dicke components. This property is essential to the analytical proof below, as it permits an explicit evaluation of the \(\rA(t)\) and \(\rADE\), from which the exact vanishing of \(\dop(t)\) follows at all times. A detailed discussion on the relation between spin-coherent states and Dicke states is given in Appendix~\ref{apdx:dicke_states}.

Following Eq.~\eqref{eq:SCS_dicke_time_evolve}, the corresponding full density matrix is given as
\begin{equation}
    \hat\rho(t)
    \!=\!
    \ket{\Psi(t)}\bra{\Psi(t)}
    \!=\!
    \sum_{k,k'=0}^{N}
    A_k A_{k'}^*
    e^{-i(E_k-E_{k'})t}
    \ket{D_k^N}\bra{D_{k'}^N}.
    \label{eq:rho_full_dicke}
\end{equation}
We next divide the full system into subsystems \(A\) and \(B\), with \(N_A\) and
\(N_B=N-N_A\) spins, respectively. The Dicke state correspondingly decomposes as  \cite{Seshadri2018}
\begin{equation}
\ket{D_k^N}
=
\sum_{\substack{
r=0,\ldots,N_A\\
s=0,\ldots,N_B
}}
F_{r,s}^{(k)}
\ket{D_r^{N_A}}_A
\otimes
\ket{D_s^{N_B}}_B ,
\label{eq:dicke_bipartition}
\end{equation}
with $r+s=k$, and the coefficient $F_{r,s}^{(k)}$ given by
\begin{equation}
\label{eq:dicke_decompose_coeff}
F_{r,s}^{(k)}
=
\sqrt{
\frac{
\binom{N_A}{r}\binom{N_B}{s}
}{
\binom{N}{k}
}
}.
\end{equation}
Here, \(r\) and \(s\) are the number of down spins in \(A\) and \(B\),
respectively. The Dicke state \(\ket{D_r^{N_A}}_A\) is an eigenstate of \(\QA\), with
\begin{equation}
    \QA\ket{D_r^{N_A}}_A
    =
    q_A(r)\ket{D_r^{N_A}}_A,
    \quad
    q_A(r)=N_A-2r .
    \label{eq:subsystem_charge_r}
\end{equation}
Substituting Eq.~\eqref{eq:dicke_bipartition} into
Eq.~\eqref{eq:rho_full_dicke} followed by tracing over \(B\), and using the orthonormality condition of the Dicke states, one gets reduced state \(\rA(t)\) as
\begin{equation}
\begin{split}
\rA(t)
&=
\sum_{r,r'=0}^{N_A}
\sum_{s=0}^{N_B}
\,
A_{r+s}\,A_{r'+s}^{*} \,
F_{r,s}^{(r+s)}\,
F_{r',s}^{(r'+s)}
\\
&\times
e^{-i(E_{r+s}-E_{r'+s})t} \,
\ket{D_{r}^{N_A}}_A
\bra{D_{r'}^{N_A}}_A .
\end{split}
\label{eq:rhoA_collective_Dicke_delta}
\end{equation}
Charge dephasing in the \(\QA\) basis keeps only terms with the same subsystem
charge. Since \(q_A(r)=N_A-2r\), this is equivalent to keeping only
\(r=r'\). Hence,
\begin{equation}
\begin{split}
\rQA(t)
&=
\sum_{r=0}^{N_A}
\sum_{s=0}^{N_B}
|A_{r+s}|^2
\left|
F_{r,s}^{(r+s)}
\right|^2
\ket{D_{r}^{N_A}}_A
\bra{D_{r}^{N_A}}_A  \\
&= \sum_{r=0}^{N_A}
    \binom{N_A}{r}
    (u)^{2(N_A-r)}
    (v)^{2r}
    \ket{D_r^{N_A}}_A
    \bra{D_r^{N_A}}_A ,
\end{split}
\label{eq:rQA_dicke}
\end{equation}
where \(u=\cos\frac{\theta}{2}\, {\rm{and}}~  v=\sin\frac{\theta}{2}\) and we obtain 
%\bka{All phases disappear because the retained terms satisfy
%\(
%E_{r+s}-E_{r+s}=0.
%\)
%resulting to  
\(\rQA(t)=\rQA(0)\).
As a result, the time dependence of \(\rA(t)\) is entirely contained in the coherences between
different subsystem charge sectors.

We now evaluate the reduced diagonal ensemble and compare it with Eq.~\eqref{eq:rQA_dicke}. Since \([\hat H,\Q]=0\), the diagonal ensemble relevant for
the charge sector decomposition can be resolved in the simultaneous
\((\hat H,\Q)\) eigenspaces.
\begin{equation}
\rDE
=
\sum_{n,q} \,
\hat{\Pi}_{n,q}\,
\hat \rho(0) \,
\hat{\Pi}_{n,q},
\label{eq:DE_HQ_resolved_collective_delta}
\end{equation}
where \(\hat{\Pi}_{n,q}\) is the projector onto the simultaneous eigenspace of
\(\hat H\) and \(\Q\) with energy \(E_n\) and total charge \(q\). In general, \( \hat{\Pi}_{n,q}
=
\sum_{\lambda}
\ket{n,q,\lambda}\bra{n,q,\lambda},\) where \(\lambda\) labels the degeneracy.
In the fully symmetric Dicke sector of the collective model,
\[
    \hat{\Pi}_{E_k,q_k}^{(j=N/2)}
    =
    \ket{D_k^N}\bra{D_k^N},
    \qquad
    q_k=N-2k .
\]
Note that in the fully symmetric sector, Dicke state \(\ket{D_k^N}\) is a simultaneous eigenstate of both \(\hat{H}\) and \(\Q\), with each Dicke label \(k\) has a unique total
charge \(q_k=N-2k\). Thus,
\begin{equation}
\rDE
=
\sum_{k=0}^{N}
|A_k|^2
\ket{D_k^N}\bra{D_k^N}.
\label{eq:DE_HQ_Dicke_collective_delta}
\end{equation}
The corresponding reduced diagonal ensemble is
\begin{equation}
\rADE
=
\Tr_B\left(\rDE\right) =
\sum_{k=0}^{N}
|A_k|^2
\Tr_B
\left(
\ket{D_k^N}\bra{D_k^N}
\right).
\label{eq:rADE_Dicke_def_delta}
\end{equation}
Using Eq.~\eqref{eq:dicke_bipartition}, this becomes
\begin{equation}
\begin{split}
\rADE
&=
\sum_{r=0}^{N_A}\;
\sum_{s=0}^{N_B}\;
|A_{r+s}|^2\,
\left|
F_{r,s}^{(r+s)}
\right|^2\,
\ket{D_r^{N_A}}_A
\bra{D_r^{N_A}}_A \\
&= \sum_{r=0}^{N_A}
\binom{N_A}{r}
\left(u\right)^{2(N_A-r)}
\left(v\right)^{2r}
\ket{D_r^{N_A}}_A
\bra{D_r^{N_A}}_A.
\end{split}
\label{eq:rADE_dicke}
\end{equation}
The coefficient above is the binomial probability distribution for finding \(r\) down spins
inside subsystem \(A\). Comparing Eq.~\eqref{eq:rQA_dicke} and
Eq.~\eqref{eq:rADE_dicke}, we obtain \(\rQA(t)=\rADE\).
Consequently,
\begin{equation}
\dop(t)
=
\rQA(t)-\rADE
=
0 ,
\label{eq:delta_zero_dicke}
\end{equation}
and hence
\(\normone{\dop(t)}=0 \) for all \(t\).

The above result is unaffected by energy degeneracies. Indeed,
\(E_k=E_l\) for \(k\neq l\) can occur,
but \(k\neq l\) implies \(q_k\neq q_l\). Thus degeneracies occur
between different total-charge sectors and are removed by the simultaneous-resolved form of diagonal ensemble in
Eq.~\eqref{eq:DE_HQ_resolved_collective_delta}. Therefore, for the collective
spin coherent calculation, \(\dop(t)=0\) exactly.

\begin{figure*}[t]
    \centering

    %-------------------- Left panel --------------------
    \begin{minipage}[t]{0.48\textwidth}
        \centering
        \includegraphics[width=\linewidth]{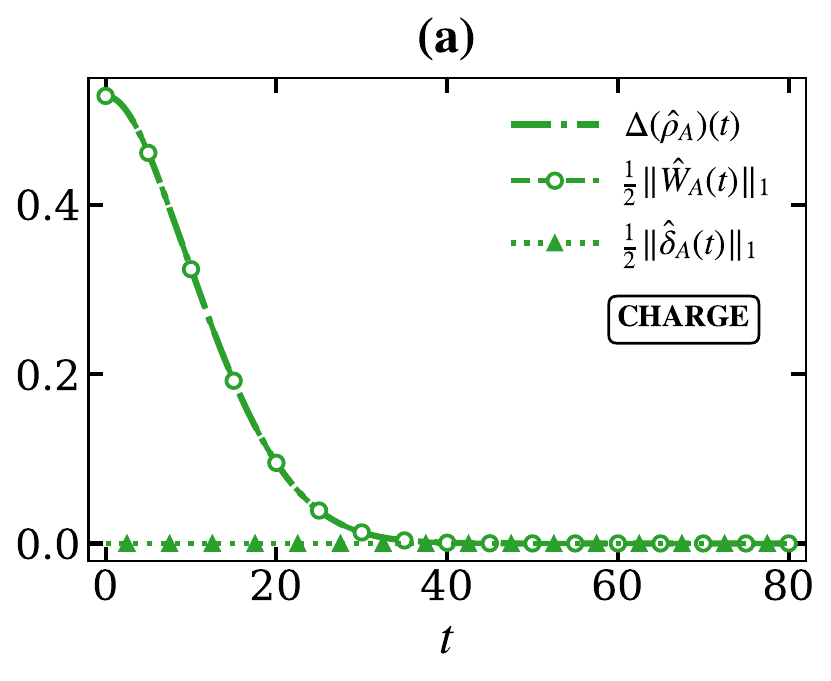}

        \vspace{1mm}
    \end{minipage}
    \hfill
    %-------------------- Right panel --------------------
    \begin{minipage}[t]{0.48\textwidth}
        \centering
        \includegraphics[width=\linewidth]{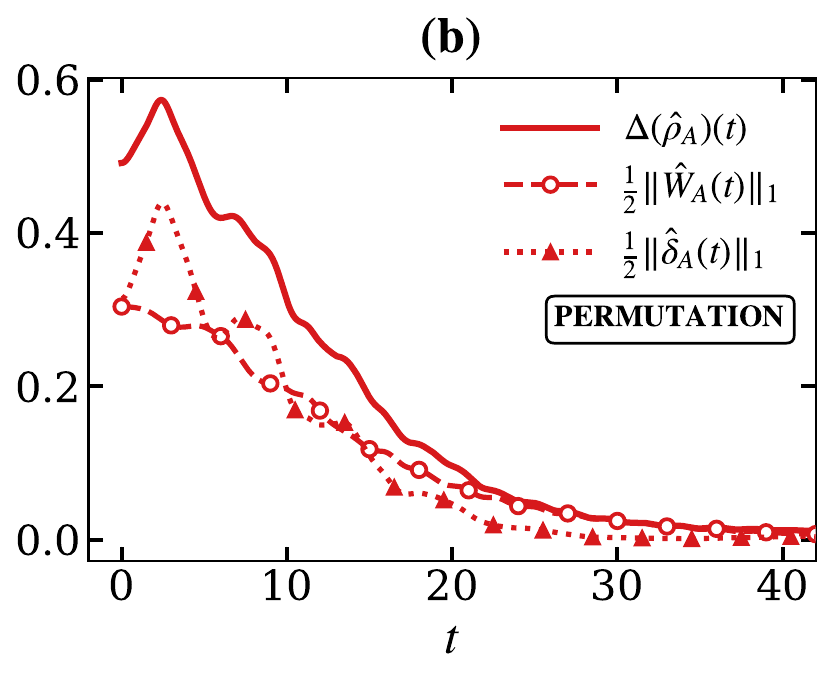}

        \vspace{1mm}
    \end{minipage}

    \caption{
    Comparison of the contributions from \(\Wop(t)\) and
    \(\dop(t)\) to the trace distance \(\Dtd(t)\) dynamics for (a) charge
    and (b) permutation symmetries.
    The charge symmetry broken initial state has \(\theta = 0.3\pi\), and is evolved with the Hamiltonian in Eq. \eqref{eq:H_U1_form} with parameters
    \(\alpha=1/N\) and \(p=\pi/2\) confirming the equivalence of \(\Dtd(t)\) and 1/2\(\normone{\Wop(t)}\). Other initial \(\theta\) values show similar results. For the permutation-symmetry
    case, initial state \(\Delta\theta = 0.3 \pi\). The plot confirms contribution from both the terms in \(\Dtd(t)\) . Other initial \(\Delta\theta\) values show similar non-trivial trend for both the quantities. 
    The dynamics is generated by the Hamiltonian in
    Eq.~\eqref{eq:H_general}, with
    \(\alpha_1=1/N\), \(\alpha_2=2/N\), and \(p=\pi/2\). The numerical results for both the figures are shown for \(N=80\), and \(N_A=2\), where the
    subsystem is drawn from the middle of the spin chain.
    }
    \label{fig:W_delta_comparison}
\end{figure*}

Fig.~(\ref{fig:W_delta_comparison}a) numerically confirms the analytical result
derived above. The trace norm of \(\dop(t)\) remains exactly zero
for all times and for all values of the initial tilt angle \(\theta\),
whereas \(\Wop(t)\) shows the full nontrivial time dependence.
This verifies that, for the collective charge-symmetric Hamiltonian and
a fully symmetric spin coherent initial state which breaks charge symmetry, the residual
deviation \(\dop(t)\) vanishes identically. Hence, the relaxation
observed in the trace distance \(\Dtd(t)\) is entirely governed by \(\Wop(t)\).

This exact cancellation is special to the collective charge-symmetric
Hamiltonian and the fully symmetric spin coherent initial state. On the other hand,
for a generic charge-conserving Hamiltonian, energy
eigenstates within a fixed charge sector are not to be angular momentum states. We show in Appendix \ref{apdx:charge_delta_weak} that in such cases,  \(\normone{\dop(t)}\) does not vanish, and can show a very weak time dependence.

%==================================

\subsection{Permutational symmetry: non--additive conserved quantity}
\label{subsec:permutation_nonadditive}

We now consider the general limit of our collective Hamiltonian in Eq.~\eqref{eq:H_general} (\(\alpha_1\neq\alpha_2\)) which only supports permutational symmetry. In this subsection, we show that the residual contribution \(\dop(t)\) is not in block diagonal form (non-zero in general) unlike the charge symmetry case, and therefore is dynamically important.

Since \([\hat H,\Jtwo]=0\), it immediately follows that \([\rDE,\Jtwo]=0 \). On the other hand, for the subsystem, one would obtain, 
\begin{equation}
    [\rADE,\JAtwo]
    =
    -2\Tr_B
    \left[
        \rDE,\JA\cdot\JB
    \right],
    \label{eq:J2_obstruction}
\end{equation}
which is generically nonzero, unlike the case for charge symmetry, as given in Eq.~\eqref{eq:rhoADE_commutes_QA}. Therefore, \(\rADE\) is not guaranteed to be block diagonal in the subsystem \(J_A^2\) spin sectors. The residual contribution in \(\Dtd(t)\) given by \(\dop(t)\) does not vanish in the permutational symmetry case, making it dynamically important similar to \(\Wop(t)\). Since \(\EA(t)\) is controlled by the symmetry-breaking coherence \(\Wop(t)\), while \(\Dtd(t)\) receives substantial contributions from \(\dop(t)\) in addition to \(\Wop(t)\), the two diagnostics need not show the same temporal behavior. We thus conclude that \(\Dtd(t)\) would not probe a clean Mpemba--like symmetry breaking relaxation dynamics in the quantum system as shown by the entanglement asymmetry \(\EA(t)\).

Fig.~(\ref{fig:W_delta_comparison}b) numerically confirms the
above discussion. For the permutational-symmetry case, the
trace distance \(\Dtd(t)\) receives contributions from both
\(\Wop(t)\) and \(\dop(t)\). Since trace norm of
\(\dop(t)\) remains finite and comparable in magnitude
to \(\Wop(t)\), the trace distance cannot be
interpreted as a direct measure of symmetry-sector coherence alone.

This explains why the trace distance \(\Dtd(t)\) need not be a good measure to probe Mpemba-like relaxation like its counterpart entanglement
asymmetry \(\EA(t)\), in the case of permutational symmetry.\\

%=================================================
% FIG. COMPARISON OF W_A(t) AND delta_A(t)
% FOR CHARGE AND PERMUTATIONAL SYMMETRIES
%=================================================

\subsection{Perturbative expansion of entanglement entropy $S\left(\rA(t)\right)$}
\label{subsec:perturbative_EA}
In this subsection we use perturbation theory to relate the entanglement asymmetry, defined in Eq.~\eqref{eq:EA_def} with \(\Wop(t)\), defined in Eq.~\eqref{eq:W_def}. While the operator \(\Wop(t)\), is the difference of two matrices, entanglement asymmetry is the difference of von Neumann entropy of the same two matrices.
% showing the non trivial relation between the two.
At a particular instant of time \(t\), we write,
\begin{equation}
    \rA(t)=\rACA(t)+\Wop(t),
    \label{eq:rhoACA_W_split}
\end{equation}
where \(\rACA(t)\) is block diagonal in the chosen subsystem symmetry sectors and \(\Wop(t)\) contains only the off-block-diagonal coherences. Let us now introduce an auxiliary parameter \(\epsilon\) such that,
\begin{equation}
    \rA(\epsilon,t)
    =
    \rACA(t)+\epsilon\,\Wop(t),
    \quad
    |\epsilon|\ll 1 .
    \label{eq:epsilon_state}
\end{equation}
Taylor expansion of  \(S\left(\rA(\epsilon,t)\right)\) about symmetry dephased reduced state \(\rACA(t)\), gives,
\begin{equation*}
S\big(\hat{\rho}_A(\epsilon,t)\big)
=
S(\rACA(t))
+ \sum_{n=1}^{\infty} \frac{\epsilon^n}{n!} \, S^{(n)}(t),
\end{equation*}
where
\begin{equation}
S^{(n)}(t)\equiv
\left.\frac{d^n}{d\epsilon^n}S\big(\hat{\rho}_A(\epsilon,t)\big)\right|_{\epsilon=0}
\end{equation} 
is the n-th order perturbative term.
The entanglement asymmetry then takes a form given by
\begin{equation}
\Delta S_A(\epsilon,t)
=
S(\rACA(t))\!-\!S(\hat{\rho}_A(\epsilon,t))
=
-\sum_{n= 1}^{\infty}\frac{\epsilon^n}{n!}\,S^{(n)}(t)
\label{eq:DeltaS_from_Sn}
\end{equation}
As shown in Appendix \ref{apdx:EA_perturbation}, the first-order term in the above expansion vanishes because \(\rACA(t)\) is block diagonal, while \(\Wop(t)\) is purely off-block-diagonal so that the leading contribution  of entanglement asymmetry is quadratic in \(\Wop(t)\), and is given by 
\begin{eqnarray}
    \EA(t)
    &=&
    \frac{1}{2}
    \sum_{i,j}
    |w_{ij}(t)|^2 \,
    \frac{\ln p_i(t)-\ln p_j(t)}
    {p_i(t)-p_j(t)} \nonumber \\
    &+& \mathcal{O}\!\left(\|\Wop(t)\|^3\right),
    \label{eq:EA_2order}
\end{eqnarray}
where, 
\begin{eqnarray}
\rACA(t)&=&\sum_i p_i(t) \ket{\chi_i}\bra{\chi_i}, \\
w_{ij}(t)&\equiv& \bra{\chi_i}\Wop(t) \ket{\chi_j}.
\end{eqnarray}
%Since the kernel in Eq.~\eqref{eq:EA_2order} is positive for positive
%\(p_i,p_j\), the leading contribution is positive and quadratic in the
%off-sector coherence \(\Wop(t)\).
%Thus, entanglement asymmetry \(\EA(t)\) takes the form after fixing \(\epsilon=1\),
%\begin{equation}
%\begin{split}
%\Delta S_A(t)
%=&\,
%\sum_{n\ge 2}^{\infty}\frac{(-1)^n}{n}
%\sum_{i_1,\dots,i_n}
%w_{i_1 i_2}(t)w_{i_2 i_3}(t)\cdots w_{i_n i_1}(t)
%\\
%&\qquad\qquad \qquad\qquad \times
%M_n(p_{i_1},\dots,p_{i_n})(t) 
%\end{split}
%\label{eq:EA_perturbative}
%\end{equation}
Thus, the perturbative expansion shows that \(\EA(t)\) is directly related to the matrix elements of \(\Wop(t)\), with coefficients
determined by the instantaneous spectrum of the block-diagonal state $\rACA(t)$. We also present a detailed derivation of general \(n\)-th order contribution in Appendix~\ref{apdx:EA_perturbation}.

Similar form of perturbative expansion of entanglement asymmetry has also been reported in a recent work \cite{zheng2026_spectralchaos}, in the context of effect of chaos in the $U(1)$ symmetry based \rm{QME}.

\section{Summary and Discussions}
\label{sec:summary}

In this work, we have obtained a general framework to understand when symmetry based diagnostics of Mpemba effect like entanglement asymmetry, and energy based diagnostics like trace distance, show similar behavior for different symmetries. 
For this, we decompose the trace distance in a general symmetry case into symmetry-coherence and residual contributions. This decomposition shows that the trace distance follows the entanglement asymmetry, and thus shows Mpemba like relxation only when the residual contribution has a sufficiently weak time dependence. For charge symmetry, we combine analytical arguments with numerical results to demonstrate that the residual contribution is either negligible or only weakly time dependent, explaining the observed correspondence between trace distance and entanglement asymmetry. In contrast, permutation symmetry exhibits a nontrivial evolution of the residual contribution, leading to a trace distance dynamics that does not show a clear Mpemba like relaxation behavior.
%differ qualitatively from those of entanglement asymmetry, and does not show a clear Mpemba like relaxation behavior. 
We thus conclude that the entanglement asymmetry is a more reliable diagnostic for the Mpemba effect. Our numerical results for collective Hamiltonians extend up to system sizes $N=80$ for both charge and permutation symmetry case, allowing us to rule out significant finite-size effects.
Our work therefore establishes that trace distance does not provide a faithful measure for detecting Mpemba-like relaxation associated with permutation symmetry, whereas entanglement asymmetry remains a reliable probe for both classes of symmetries.

It would be interesting to test the framework developed in our work in other many-body interacting permutationally symmetric systems, as well as extending our analysis for more general classes of symmetries. Furthermore, the Mpemba effect has recently been investigated in periodically driven closed systems with charge symmetry. An interesting future direction would be to compare the dynamics of trace distance and entanglement asymmetry in such driven systems, and to investigate whether the conditions established for undriven systems, as done in this work, can be generalized to periodically driven settings.

\section{Acknowledgments}
BKA acknowledges the CRG grant No. CRG/2023/003377 from ANRF, Government of India. UD  also acknowledges  support from ANRF through grant No. SPG/2022/000708.

%%%%%%%%%%%%%%%%%%%%%%%%%%%%%%%%%%%%%%%%%%%%%%%%%%%%%%%%%%%%%%%
%%%%%%%%%%%%%%%%%%%%%%%%%%%%%%%%%%%%%%%%%%%%%%%%%%%%%%%%%%%%%%%

\newpage

\appendix

\section{Relation between $\hat{J}^2$ spin operator and permutation symmetry}
\label{apdx:J2_permutation}
In this section, we provide a relation between the total angular momentum operator \(\hat{J}^2\) and the permutation symmetry discussed in the main text. For a system of \(N\) spin-\(\frac12\) particles, the collective spin operator is given by
\(
    \hat{\vec{J}}
    =
    1/2
    \sum_{i=1}^{N}
    \hat{\vec{\sigma}}_{i},
\) where \(\hat{\vec{\sigma}}_{i} = (\hat{\sigma}_{i}^x,\hat{\sigma}_{i}^y,\hat{\sigma}_{i}^z)\)
is the vector of Pauli operators acting on the \(i\)-th spin.
Therefore,
\begin{equation}
    \hat{J}^2
    =
    \frac{1}{4}
    \sum_{i,j=1}^{N}
    \hat{\vec{\sigma}}_{i}
    \cdot
    \hat{\vec{\sigma}}_{j} .
\end{equation}
Separating the diagonal and off-diagonal terms gives
\begin{equation}
    \hat{J}^2
    =
    \frac{3N}{4}\hat{\mathbb{I}}
    +
    \frac{1}{4}
    \sum_{i\neq j}
    \hat{\vec{\sigma}}_{i}
    \cdot
    \hat{\vec{\sigma}}_{j} .
\end{equation}
For two spin-\(\frac12\) particles, the permutation (swap) operator is \cite{Dirac1935},
\begin{equation}
    \hat{P}_{ij}
    =
    \frac{1}{2}
    \left(
        \hat{\mathbb{I}}
        +
        \hat{\vec{\sigma}}_{i}
        \cdot
        \hat{\vec{\sigma}}_{j}
    \right).
\end{equation}
Hence,
\(
    \hat{\vec{\sigma}}_{i}
    \cdot
    \hat{\vec{\sigma}}_{j}
    =
    2\hat{P}_{ij}
    -
    \hat{\mathbb{I}} \). 
Substituting this into the expression for \(\hat{J}^2\), one obtains
\begin{equation}
    \hat{J}^2 =
    \sum_{i<j}\hat{P}_{ij} \,-\,
    \frac{N(N-4)}{4}\hat{\mathbb{I}}.
\label{eq:J2_permutation_relation}
\end{equation}
Eq.~\eqref{eq:J2_permutation_relation} shows that \(\hat{J}^2\) is directly
related to the sum of all pairwise permutation operators. Therefore, the eigenvalue
of \(\hat{J}^2\) characterizes how the many-spin state transforms under particle
permutations. In particular, the fully permutation-symmetric subspace corresponds
to the maximum total-spin sector \(j=N/2\). Any finite weight in sectors
\(j<N/2\) signals that the state is no longer fully permutation symmetric, thus indicating the permutation symmetry breaking in the given quantum state. This relation in Eq.~\eqref{eq:J2_permutation_relation} between the total spin operator \(\Jtwo\) and the permutation symmetry helps us to probe the permutation symmetry breaking dynamics through irreducible representation of possible spin sectors \(j\).\\

\section{Dicke states and Spin coherent states}
\label{apdx:dicke_states}
In this section, we discuss relation between Dicke states and the tilted ferromagnetic state, as given in Eq.~(\ref{eq:charge_initial_state}). Such a state is used in our work for studying charge-symmetry breaking dynamics. The Dicke states form the natural basis of the permutation-symmetric
subspace of an \(N\)-spin system \cite{Dicke1954,Seshadri2018,Ieminietal}. It is defined as
\begin{equation}
    \ket{D_k^N}
    =
    \frac{1}{\sqrt{\binom{N}{k}}}
    \sum_{\mu:\,w(\mu)=k}
    \ket{\mu},
    \label{eq:Dicke_def_delta}
\end{equation}
where \(w(\mu)=k\) denotes the Hamming weight of the computational basis
state \(\ket{\mu}\). In the present convention, \(k\) is the number of down
spins (\(-z\) direction) in a total \(N\)-spin system. 
In the fully symmetric sector (\(j=N/2\)), it is convenient to use $\ket{j,m}$ basis so that 
\[
    \ket{D_k^N}\equiv \ket{j,m_k},
    \qquad
    m_k=j-k,
\]
where, as before, \(k\) is the number of down (-$z$) spins. These states satisfy the standard
angular-momentum eigenvalue relations
\[
    \hat J^2\ket{j,m_k}=j(j+1)\ket{j,m_k},
    \quad
    \hat J_z\ket{j,m_k}=m_k\ket{j,m_k}.
\]
For the collective charge-symmetric Hamiltonian, given in Eq.~\eqref{eq:H_U1_form}, one therefore obtains
\[
    \hat H\ket{D_k^N}
    =
    E_k\ket{D_k^N},
\]
with
\[
    E_k = \bra{j,m_k} \,\hat H\, \ket{j,m_k}
    =
    \alpha\left[j(j+1)-m_k^2\right]+pm_k.
\]
Thus the expression for \(E_k\) follows directly from the angular-momentum
eigenvalue equations.

We now derive the expansion coefficient \(A_k\) of a spin coherent state in
the Dicke basis, see Eq.~\eqref{eq:SCS_dicke_expansion} of the main text. A spin coherent state can be written as a product of identical
single-spin states:
\begin{equation}
\ket{\Psi_0}
=
\left(
\cos\frac{\theta}{2}\ket{0}
+
e^{i\phi}\sin\frac{\theta}{2}\ket{1}
\right)^{\otimes N} =
\sum_{k=0}^{N}
A_k\ket{D_k^N}.
\label{eq:appendix_SCS_product}
\end{equation}
Let, \(
u=\cos\frac{\theta}{2}\;,\;
v=e^{i\phi}\sin\frac{\theta}{2}\). Then
\begin{equation}
\ket{\Psi_0}
=
\left(
u\ket{0}
+
v\ket{1}
\right)^{\otimes N}.
\end{equation}
Expanding this tensor product binomially gives
\begin{equation}
\begin{split}
\ket{\Psi_0}
&=
u^N\ket{00\cdots00}
\;
+ \;
u^{N-1}v
\sum_{i_1}
\ket{1_{i_1}}
\\
&\quad
+
u^{N-2}v^2
\sum_{i_1<i_2}
\ket{1_{i_1}1_{i_2}}
\\
&\quad
+\cdots
\\
&\quad
+
u^{N-k}v^k
\sum_{i_1<i_2<\cdots<i_k}
\ket{1_{i_1}1_{i_2}\cdots1_{i_k}}
\\
&\quad
+\cdots
+
v^N\ket{11\cdots1}.
\end{split}
\label{eq:appendix_SCS_binomial_expansion}
\end{equation}
Here \(\ket{1_{i_1}1_{i_2}\cdots1_{i_k}}\)
denotes the computational-basis state in which the spins at sites
\(i_1,i_2,\ldots,i_k\) are  in \(\ket{1}\) state, and all remaining spins are in $|0\rangle$ state. Equivalently, the expansion can be written compactly as
\begin{equation}
\ket{\Psi_0}
=
\sum_{k=0}^{N}
u^{N-k}v^k
\sum_{1\leq i_1<i_2<\cdots<i_k\leq N}
\ket{1_{i_1}1_{i_2}\cdots1_{i_k}},
\label{eq:appendix_SCS_grouped_k}
\end{equation}
with the summation over all distinct choices of \(k\) down-spin positions among the \(N\) sites. The ordering condition
\(1\leq i_1<i_2<\cdots<i_k\leq N\) avoids double counting. The normalized Dicke state with \(k\) down spins is then given by \cite{Seshadri2018},
\begin{equation}
\ket{D_k^N}
=
\frac{1}{\sqrt{\binom{N}{k}}}
\sum_{1\leq i_1<i_2<\cdots<i_k\leq N}
\ket{1_{i_1}1_{i_2}\cdots1_{i_k}},
\label{eq:appendix_Dicke_definition}
\end{equation}
where, \(\binom{N}{k} =  \frac{N!}{k!(N-k)!}\) is the binomial coefficient. Therefore,
\begin{equation}
\sum_{1\leq i_1<i_2<\cdots<i_k\leq N}
\ket{1_{i_1}1_{i_2}\cdots1_{i_k}}
=
\sqrt{\binom{N}{k}}\ket{D_k^N}.
\end{equation}
Substituting this into Eq.~\eqref{eq:appendix_SCS_grouped_k}, we obtain
\begin{equation}
\ket{\Psi_0}
=
\sum_{k=0}^{N}
\sqrt{\binom{N}{k}}\,
u^{N-k}v^k
\ket{D_k^N}.
\end{equation}
Using the definitions of \(u\) and \(v\), and comparing with Eq. \ref{eq:appendix_SCS_product}, we get
\begin{equation}
A_k
=
\sqrt{\binom{N}{k}}
\left(\cos\frac{\theta}{2}\right)^{N-k}
\left(e^{i\phi}\sin\frac{\theta}{2}\right)^k .
\label{eq:appendix_Ak}
\end{equation}
One can also check the normalization as follows:
\begin{equation}
\begin{split}
\sum_{k=0}^{N}|A_k|^2
&=
\sum_{k=0}^{N}
\binom{N}{k}
\left(\cos^2\frac{\theta}{2}\right)^{N-k}
\left(\sin^2\frac{\theta}{2}\right)^k
\\
&=
\left(
\cos^2\frac{\theta}{2}
+
\sin^2\frac{\theta}{2}
\right)^N
=
1 .
\end{split}
\end{equation}
Finally, since each Dicke state \(\ket{D_k^N}\) is an eigenstate of
\(\hat H\), the time-evolution operator acts as
\begin{equation}
\begin{split}
\ket{\Psi(t)}
&=
e^{-i\hat Ht}\ket{\Psi_0}
\\
&=
\sum_{k=0}^{N}
A_k e^{-iE_kt}\ket{D_k^N}.
\end{split}
\label{eq:appendix_SCS_time_evolution}
\end{equation}
Thus, for the collective charge-symmetric Hamiltonian, the dynamics does not
mix different Dicke components. It only attaches the phase \(e^{-iE_kt}\) to
each Dicke component.

%===========================================

\section{Weak time dependence of \(\dop(t)\) in the generic  charge-symmetric model}
\label{apdx:charge_delta_weak}

In this section, we derive the explicit matrix elements of \(\dop(t)\) for a generic non-integrable Hamiltonian having charge symmetry. Since
\([\hat H,\Q]=0\), we choose a simultaneous eigenbasis of
\(\hat H\) and \(\Q\), denoted by \(|n,q\rangle\), where \(q\) labels
the total charge sector and \(n\) labels energy eigenstates inside
that sector \cite{charge_YiHan_et.al}. For simplicity, we assume no degeneracy. These states satisfy the relations

\begin{equation}
    \hat{H}|n,q\rangle=E_n^q|n,q\rangle,
    \qquad
    \Q|n,q\rangle=q|n,q\rangle .
    \label{eq:energy_charge_basis}
\end{equation}
The initial state can be expanded as
\begin{equation}
|\Psi_0\rangle
=
\sum_{n,q} c_{nq}|n,q\rangle ,
\qquad
c_{nq}=\langle n,q|\Psi_0\rangle ,
\end{equation}
and hence
\begin{equation}
    |\Psi(t)\rangle
    =
    \sum_{n,q}
    c_{nq}e^{-iE_n^q t}|n,q\rangle.
\end{equation}
Therefore, the full density matrix can be written as
\begin{equation}
    \hat{\rho}(t)
    =
    \sum_{q,q'}
    \sum_{n,m}
    c_{nq}\,c^*_{mq'}
    e^{-i(E_n^q-E_m^{q'})t}
    |n,q\rangle\langle m,q'| \,.
    \label{eq:rho_full_U1}
\end{equation}
This expression contains coherences between different total-charge sectors \((q\neq q')\), as well as coherences between different energy eigenstates within the same charge sector \((q=q', n\neq m)\).

Since \(\ket{n,q}\) belongs to the total-charge sector \(q\), it can be expanded in the product eigenbasis of \(\QA\) and \(\QB\). We choose subsystem charge bases as 
\begin{equation}
    \QA|a,r\rangle_A=r|a,r\rangle_A,
    \qquad
    \QB|b,s\rangle_B=s|b,s\rangle_B, 
\end{equation}
where charges \(r \in q_A\), \(s \in q_B\) and \(a,b\) corresponds to the energy indices for the respective subsystems. This product state
\(\ket{a,r}_A \otimes\ket{b,s}_B\) is also an eigenstate of the total charge \(\hat Q\). Since \(q=r+s\) due to additive nature of the charge symmetry as discussed in Sec. \ref{subsec:additive_nonadditive},  total-charge eigenstate can be written as
\begin{equation}
    |n,q\rangle
    =
    \sum_r\sum_{a,b}
    C^{(n,q)}_{ar;b(q-r)}
    |a,r\rangle_A \otimes |b,q-r\rangle_B \,.
    \label{eq:charge_expansion}
\end{equation}
The coefficient \(C^{(n,q)}_{ar;\,b(q-r)}
={}_A\!\bra{a,r}\otimes{}_B\!\bra{b,q-r}\,n,q\rangle\) is nonzero only when the subsystem charges \(r,s\) add to
the total charge \(q\). Thus, only
those product states satisfying the charge constraint \(r+s=q\) can appear.\\

We now compute the matrix elements of the reduced density matrix in the subsystem charge basis. From Eq.~\eqref{eq:rho_full_U1},
\begin{equation}
\begin{split}
\langle a,r|\,\hat \rho_A(t)\,|a',r'\rangle
&=
\sum_{q,q'}
\sum_{n,m}
c_{nq}c^{*}_{mq'}
e^{-i(E_n^q-E_m^{q'})t}
\\
&\quad \times
\langle a,r|
\Tr_B\!\left(|n,q\rangle\langle m,q'|\right)
|a',r'\rangle .
\end{split}
\end{equation}
Using the expansion in Eq.~\eqref{eq:charge_expansion}, the partial
trace gives
\begin{equation}
\begin{split}
&\langle a,r|
\Tr_B\!\left(|n,q\rangle\langle m,q'|\right)
|a',r'\rangle
\\
&=
\delta_{q-r,\,q'-r'}
\sum_b
C^{(n,q)}_{ar;\,b(q-r)}
\left[
C^{(m,q')}_{a'r';\,b(q'-r')}
\right]^* .
\end{split}
\label{eq:partial_trace_charge_selection}
\end{equation}
The Kronecker delta expresses the charge-selection rule
\begin{equation}
q-r=q'-r',
\qquad\text{or equivalently}\qquad
q-q'=r-r' .
\label{eq:charge_selection_rule}
\end{equation}
Thus, a reduced density-matrix element connecting subsystem charge
sectors \(r\) and \(r'\) can arise only from full-system coherences
connecting total charge sectors \(q\) and \(q'\) satisfying
Eq.~\eqref{eq:charge_selection_rule}.

Now we compute matrix elements of \(\rQA(t)\), keeping only the block-diagonal elements of \(\rA(t)\) in the \(\QA\) basis, namely
terms with \(r=r'\). Using the selection rule in Eq.~\eqref{eq:charge_selection_rule},
this immediately implies \(q=q'\)
Therefore, charge dephasing on subsystem \(A\) removes all
coherences between different total-charge sectors and hence,

%======================================
\begin{widetext}
\begin{align}
    \langle a,r| \,
    \rQA(t) \,
    |a',r\rangle
    =
    \sum_q\sum_{n,m} \,
    c_{nq}c^*_{mq}\,
    e^{-i(E_n^q-E_m^q)t} \,
    \sum_b \,
    C^{(n,q)}_{ar;b(q-r)}
    \left[
        C^{(m,q)}_{a'r;b(q-r)}
    \right]^*  \,.
    \label{eq:rhoAQ_U1_matrix}
\end{align}
\end{widetext}

Let
$
\hat K^{(q,r)}_{nm}
=
\mathcal{\hat P}_r^A \;
\text{Tr}_B\left(\ket{n,q}\bra{m,q}\right) \;
\mathcal{\hat P}_r^A,$
where \(\hat{\mathcal{P}}^{A}_{r}\) is the projector corresponding to subsystem charges \(r \in Q_A\).
Its matrix elements are
\[
{}_A\bra{a,r}\hat K^{(q,r)}_{nm}\ket{a',r}_A
=
\sum_b
C^{(n,q)}_{ar;\,b(q-r)}
\left(C^{(m,q)}_{a'r;\,b(q-r)}\right)^*
\]
Therefore, \(\rQA(t)\) in matrix form can be written as,
\begin{equation} \label{eq:Pi_rhoA_opform}
\rQA(t)
=
\sum_r\sum_q\sum_{n,m}\;
c_{nq}c^*_{mq}\,
e^{-i(E_n^q-E_m^q)\,t}\,
\hat K^{(q,r)}_{nm}
\end{equation}
%The reduced diagonal ensemble \(\rADE\) corresponds to the \(n=m\) contribution of Eq.~\eqref{eq:rhoAQ_U1_matrix}. Hence, in the charge
%symmetry case,
%===================================================
Since diagonal ensemble \(\rDE\) is defined as the infinite-time average of the full density matrix \cite{diag_ensemble2017}, the diagonal ensemble can be written as
\begin{equation}\label{eq:rho_DE_define}
\hat \rho_{\mathrm{DE}}
=
\sum_q\sum_n
|c_{nq}|^2
\ket{n,q}\bra{n,q}.   
\end{equation}
The reduced diagonal ensemble is
\begin{equation}\label{eq:rhoDE_A_define}
    \hat \rho^A_{\mathrm{DE}}
=
\text{Tr}_B(\hat\rho_{\mathrm{DE}})
=
\sum_{n,q}
|c_{nq}|^2 \;
\text{Tr}_B\left(\ket{n,q}\bra{n,q}\right).
\end{equation}
Thus, reduced diagonal ensemble \(\rADE\) is obtained by keeping only the
diagonal energy contributions \(n=m\). Its matrix elements are
\begin{equation}
\begin{split}
&\langle a,r|\,\rADE\,|a',r\rangle
\\
&=
\sum_q
\sum_n
|c_{nq}|^2
\sum_b
C^{(n,q)}_{ar;\,b(q-r)}
\left[
C^{(n,q)}_{a'r;\,b(q-r)}
\right]^* .
\end{split}
\label{eq:rhoADE_U1_matrix}
\end{equation}

Using \(\hat K^{(q,r)}_{nn}\), \(\rADE\) in matrix form becomes,
\begin{equation} \label{eq:rhoDE_A_opform}
\hat\rho^A_{\mathrm{DE}}
=
\sum_r\sum_q\sum_n
|c_{nq}|^2
\hat K^{(q,r)}_{nn} 
\end{equation}

Subtracting Eq.~\eqref{eq:rhoADE_U1_matrix} from
Eq.~\eqref{eq:rhoAQ_U1_matrix}, the diagonal terms \(n=m\)
cancel exactly. Therefore,

%======================================
\begin{widetext}
\begin{align}
    \langle a,r| \,
    \dop(t) \,
    |a',r\rangle
    =
    \sum_q
    \sum_{\substack{n,m\\n\neq m}} \,
    c_{nq}c^*_{mq} \,
    e^{-i(E_n^q-E_m^q)t}\,
    \sum_b\,
    C^{(n,q)}_{ar;b(q-r)}
    \left[
        C^{(m,q)}_{a'r;b(q-r)}
    \right]^* \,.
    \label{eq:delta_U1_matrix}
\end{align}
\end{widetext}

In operator form,
\begin{equation} \label{eq:delta_t_opform}
\dop(t)
=
\sum_r\sum_q\sum_{n\neq m} \;
c_{nq}c^*_{mq}\,
e^{-i(E_n^q-E_m^q)t}\,
\hat K^{(q,r)}_{nm}
\end{equation}

Thus, in the
charge-symmetry case, \(\dop(t)\) contains only same charge-sector, energy-off-diagonal coherences. The diagonal energy contribution cancels exactly against \(\rADE\), and the inter-subsystem charge sector coherences are removed by \(\rQA\). This strong restriction is the reason \(\dop(t)\) acts only as a residual block-diagonal fluctuation in the charge symmetry case.

In what follows, we justify the weak time dependence of \(\dop(t)\) when compared to \(\Wop(t)\) in the generic case of charge symmetry.

Using Eq.~\eqref{eq:delta_U1_matrix}, we can write the block-diagonal operator \(\dop(t)\) as
\begin{equation}
    \dop(t)
    =
    \bigoplus_r \hat{\delta}_{A,r}(t),
    \label{eq:delta_block_decomp}
\end{equation}
where
\begin{equation}
    \hat{\delta}_{A,r}(t)
    =
    \sum_q\sum_{n\neq m}
    c_{nq}c^*_{mq}
    e^{-i(E_n^q-E_m^q)t}
    \hat{K}^{(q,r)}_{nm},
    \label{eq:delta_block}
\end{equation}
with \(
    \hat{K}^{(q,r)}_{nm}
    =
    \hat{\mathcal{P}}^{A}_{r} \,\left(
    \Tr_B
    \left(
        |n,q\rangle\langle m,q|
    \right)\,\right)
    \hat{\mathcal{P}}^{A}_{r}\),
where \(\hat{\mathcal{P}}^{A}_{r}\) is the projector corresponding to subsystem charges \(r \in q_A\). All the time dependence appears through phases of the form
\(e^{-i(E_n^q-E_m^q)t}\),
while the amplitudes \(c_{nq}c^*_{mq}\hat{K}^{(q,r)}_{nm}\) are time independent. Thus, each block \(\hat{\delta}_{A,r}(t)\) can be interpreted as a sum of fixed matrices rotating with many different frequencies. The trace norm can change significantly only if these oscillating terms strongly reorganize the singular values or eigenvalue spectrum of \(\hat{\delta}_{A}(t)\). In the charge symmetry case, this reorganization is strongly constrained by the block structure and by dephasing among many frequencies. For many incommensurate frequencies, these cross terms dephase strongly and give only small residual oscillations compared to the overall temporal behaviour.\\

In order to check for this weak time dependence of \(\dop(t)\) in a generic system, the Hamiltonian we used is a chaotic short-ranged, charge-symmetric next-nearest-neighbour XXZ  spin Hamiltonian,
\begin{eqnarray}
\hat H_{XXZ}
&=\;
J_1 \sum_{i=1}^{N-1}
\left(
\hat\sigma_i^x \hat\sigma_{i+1}^x
+
\hat\sigma_i^y \hat\sigma_{i+1}^y
+
\Delta_1 \, \hat\sigma_i^z \hat\sigma_{i+1}^z
\right) \nonumber \\
&\!\!\!\!\!\!\!\!\!\!\!\!\!\!\!\!\!\!\!\!+
J_2 \sum_{i=1}^{N-2}
\left(
\hat\sigma_i^x \hat\sigma_{i+2}^x
+
\hat\sigma_i^y \hat\sigma_{i+2}^y
+
\Delta_2 \,\hat\sigma_i^z \hat\sigma_{i+2}^z
\right),
\label{eq:H_XXZ}
\end{eqnarray}
with next-nearest-neighbor interaction which makes it non-integrable and was studied in previous literature \cite{noGlobal}.

%==========================================
% FIG 6 - MODIFIED TO 4 SUBPLOTS STRUCTURE
%============================================================
% FOUR-PANEL FIGURE FOR XXZ RESULTS
%============================================================

\begin{figure*}[t]
    \centering

    %==================== Top row: (a) and (b) ====================
    \begin{minipage}[t]{0.495\textwidth}
        \centering
        \includegraphics[
            width=\linewidth
        ]{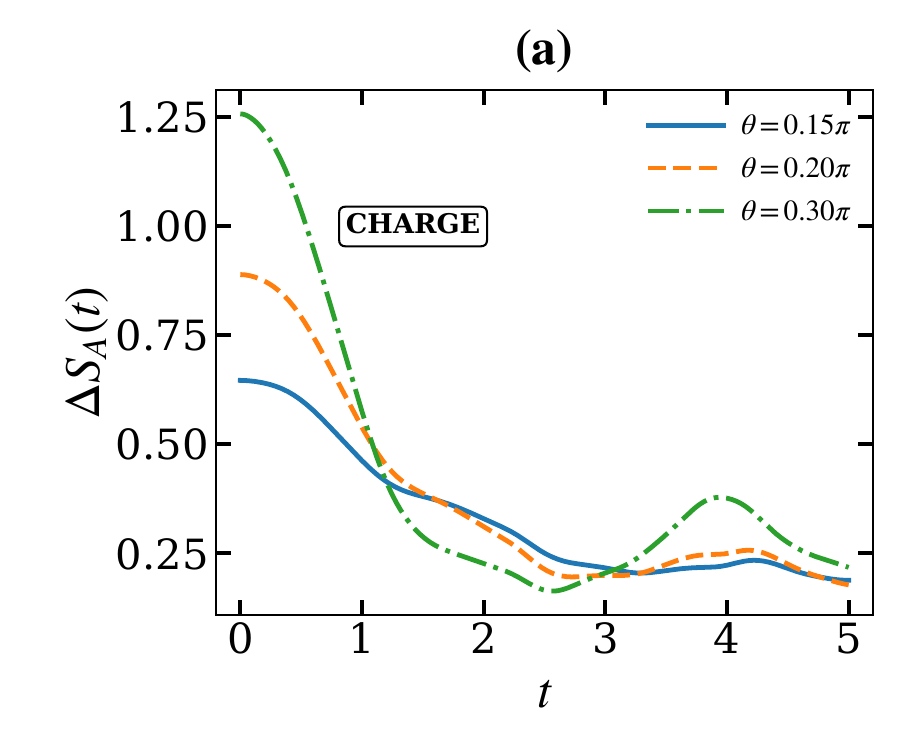}
    \end{minipage}
    \hfill
    \begin{minipage}[t]{0.495\textwidth}
        \centering
        \includegraphics[
            width=\linewidth
        ]{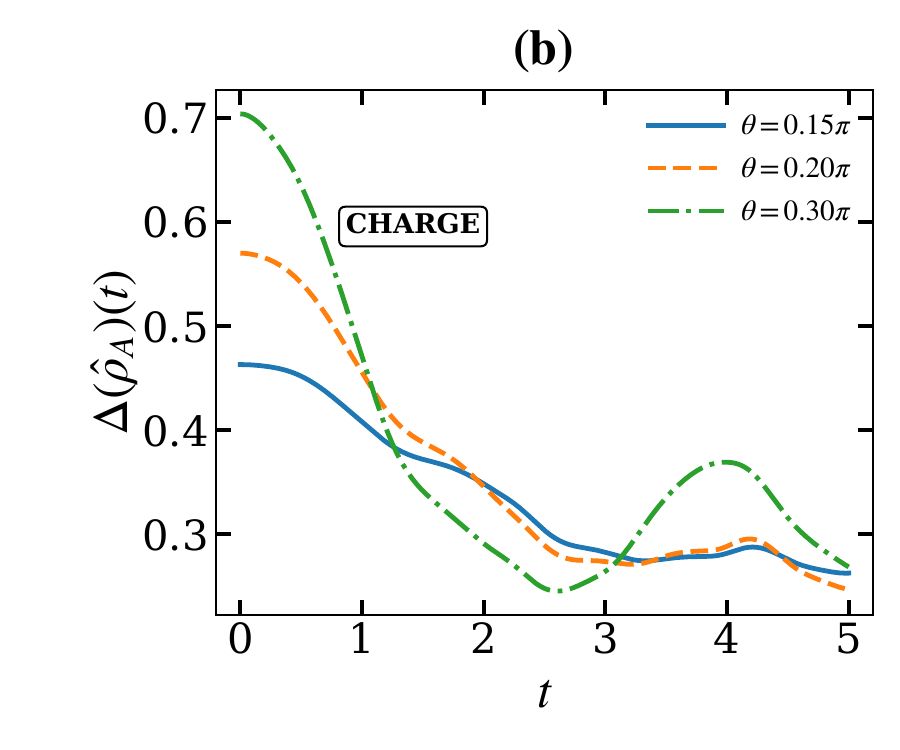}
    \end{minipage}

    \vspace{3mm}

    %==================== Bottom row: (c) and (d) ====================
    \begin{minipage}[t]{0.495\textwidth}
        \centering
        \includegraphics[
            width=\linewidth
        ]{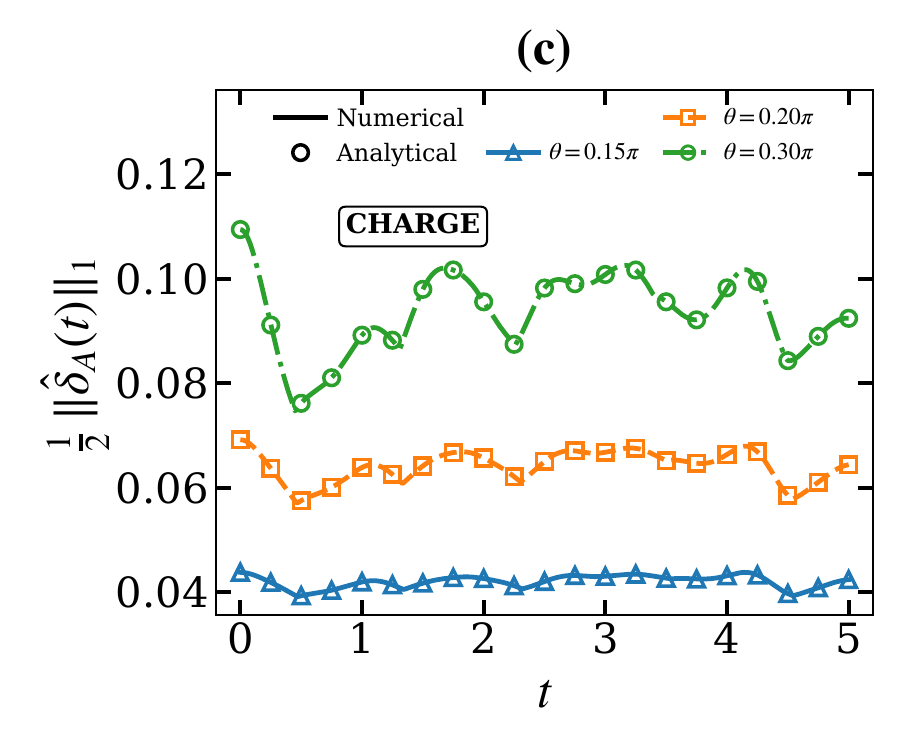}
    \end{minipage}
    \hfill
    \begin{minipage}[t]{0.495\textwidth}
        \centering
        \includegraphics[
            width=\linewidth
        ]{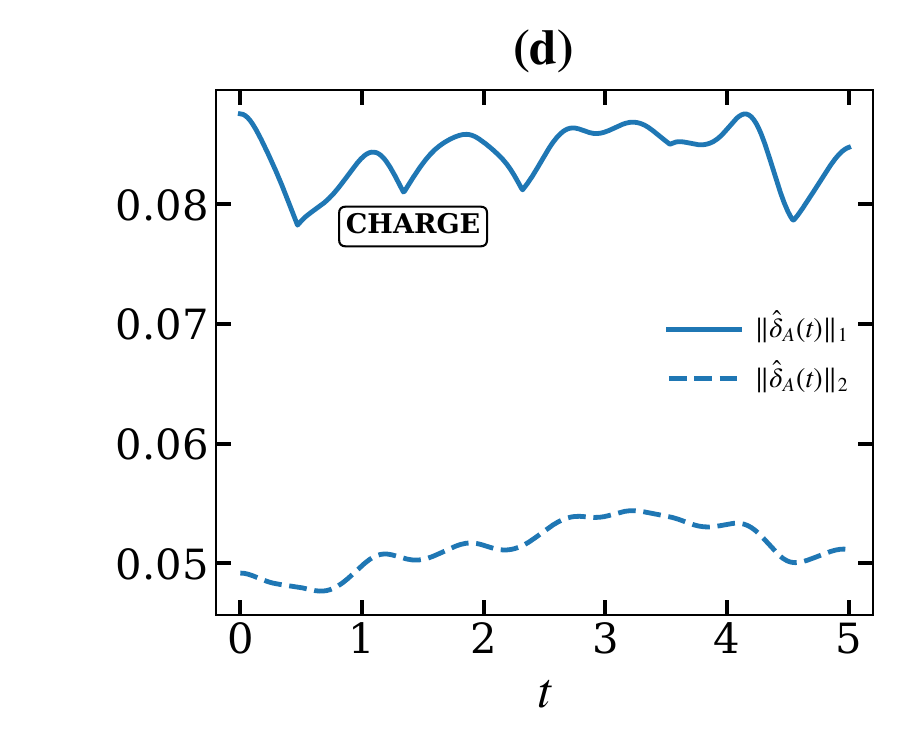}
    \end{minipage}

    \caption{
    Numerical and analytical results for the charge-symmetric
    XXZ Hamiltonian in Eq.~\eqref{eq:H_XXZ}.
    \textbf{(a)} Time evolution of the entanglement asymmetry
    \(\Delta S_A(t)\) for different initial tilt angles
    \(\theta\).
    \textbf{(b)} Corresponding time evolution of the trace
    distance
    \(\Delta(\hat{\rho}_A)(t)\) from the reduced diagonal
    ensemble.
    \textbf{(c)} Time evolution of trace norm
    \(\|\dop(t)\|_1\) for the same set of initial
    tilt angles. The lines represent the direct numerical
    results, while the symbols represent the analytical
    results obtained from Eq.~\eqref{eq:delta_U1_matrix}.
    \textbf{(d)} Comparison of the trace norm
    \(\|\dop(t)\|_1\), the Hilbert--Schmidt norm
    \(\|\dop(t)\|_2\), for the fixed initial tilt angle
    \(\theta=0.15\pi\).
    All results correspond to system size \(N=10\),
    subsystem size \(N_A=5\), and Hamiltonian parameters
    \(J_1=J_2=1\) and
    \(\Delta_1=\Delta_2=0.5\) as discussed previously in the supplementary material of Ref.~\cite{noGlobal}.
    }
    \label{fig:delta_XXZ_HS_combined}
\end{figure*}
%=========================================

From Fig.~(\ref{fig:delta_XXZ_HS_combined}a) and Fig.~(\ref{fig:delta_XXZ_HS_combined}b), we observe that the trace distance \(\Dtd(t)\) qualitatively reproduces the temporal behavior of the entanglement asymmetry \(\EA(t)\) for the generic XXZ charge-symmetric Hamiltonian equation~\eqref{eq:H_XXZ}, similar to collective charge-symmetric Hamiltonian case discussed in Sec.~(\ref{subsec:delta_exact_collective_charge})\\

Fig.~(\ref{fig:delta_XXZ_HS_combined}c) shows that \(\dop(t)\) is no
longer exactly zero, but has very weak time dependence compared with \(\Wop(t)\) for the XXZ charge symmetric Hamiltonian Eq.~\eqref{eq:H_XXZ}.
This is expected because the energy eigenstates within a fixed charge
sector are not restricted to the collective Dicke-state structure.
Therefore, same-charge, energy-off-diagonal coherences can survive after
subsystem charge dephasing. Nevertheless, the trace norm of
\(\dop(t)\) remains small and only weakly time
dependent for all the tilt angles \(\theta\) considered. Fig. (\ref{fig:delta_XXZ_HS_combined}b) also verifies the validity of the analytical expression in
Eq.~\eqref{eq:delta_U1_matrix} by comparing the direct numerical
evaluation of \(\dop(t)\) with the analytical form in simultaneous charge eigenbasis.\\

We further confirmed that the Hilbert--Schmidt norm defined as  $
    \normtwo{\dop(t)}
    =
    \sqrt{\Tr[\,\dop^\dagger(t)\dop(t)\,] }$ \cite{nielsen2010quantum} also shows weak time
    dependence, see Fig. (\ref{fig:delta_XXZ_HS_combined}~d), pointing to Mpemba like relaxation behavior in other distance based measures as well. 
    %\tam{Also, rank of the matrix \(\dop(t)\) is almost stationary in time, which indicates that the oscillatingintra-charge sector coherences do not strongly reorganize theeigenvalue spectrum of \(\dop(t)\).}

\section{Perturbative expansion of entanglement asymmetry}
\label{apdx:EA_perturbation}

In this section, we derive the perturbative relation between the
symmetry-breaking coherence operator \(\Wop(t)\) and the entanglement
asymmetry \(\EA(t)\). For a fixed time \(t\), we write the reduced state as
\begin{equation}
    \rA(t)
    =
    \rACA(t)+\Wop(t),
    \label{eq:apdx_rho_split}
\end{equation}
where
\(\rACA(t)=\Phi_{\hat{\mathcal{C}}_A}\!\left\{\rA(t)\right\}\) can be interpreted as symmetry-dephasing map acting on \(\rA(t)\). Therefore, \(\Wop(t)\) contains only the off-block-diagonal coherences between different
subsystem symmetry sectors. Since \(\Phi_{\hat{\mathcal{C}}_A}\!\left\{.\right\}\) is a linear, completely positive, trace preserving map, we get
\begin{equation}
    \Tr[\rACA(t)]=1,
    \qquad
    \Tr[\Wop(t)]=0.
    \label{eq:apdx_trace_conditions}
\end{equation}

Let us introduce an auxiliary parameter
\(\epsilon\) and define
\begin{equation}
    \hat \rho_{A}(\epsilon,t)
    =
    \rACA(t)+\epsilon\,\Wop(t),
    \qquad
    |\epsilon|\ll 1 .
    \label{eq:apdx_rho_epsilon}
\end{equation}
The actual reduced density matrix is recovered at \(\epsilon=1\). Expanding the von Neumann
entropy of \( \hat\rho_{A} (\epsilon,t)\) about $\epsilon=0$, we get 
%\(S(\hat{\rho}_{A,\epsilon})
%    =
 %   -\Tr\!\left[
 %       \hat{\rho}_{A,\epsilon} \ln \hat{\rho}_{A,\epsilon}
 %   \right]\) expanding around \(\epsilon=0\),
\begin{equation}
    S\big(\hat{\rho}_{A}(\epsilon,t)\big)
    =
    S\big(\rACA(t)\big)
    +
    \sum_{n\geq 1}
    \frac{\epsilon^n}{n!}S^{(n)},
    \label{eq:apdx_entropy_expansion}
\end{equation}
where
\begin{equation}
    S^{(n)}
    =
    \left.
    \frac{d^n}{d\epsilon^n} \big(
    S\big(\hat{\rho}_{A}(\epsilon,t)\big) \big)
    \right|_{\epsilon=0}.
\end{equation}
Therefore,
\begin{equation}
    \EA(\epsilon,t)
    =
    S(\rACA(t))-S(\hat{\rho}_{A}(\epsilon,t))
    =
    -
    \sum_{n\geq 1}
    \frac{\epsilon^n}{n!}S^{(n)}.
    \label{eq:apdx_EA_from_Sn}
\end{equation}

%\subsection{Integral representation of the logarithm of an operator}

In order to calculate the derivative of  logarithm of an operator, we use the standard integral representation of the logarithm of a positive
operator defined below: \cite{log_integral_form},
\begin{equation}
    \ln \hat{X}
    =
    \int_0^\infty d\beta
    \left[
        \frac{1}{1+\beta}\id
        -
        \frac{1}{\hat{X}+\beta\,}
    \right].
    \label{eq:apdx_log_integral}
\end{equation}
Defining
\begin{equation}
    \hat{G}_\epsilon(\beta,t)
    =
    \left(
        \hat{\rho}_{A}(\epsilon,t)+\beta\,\id
    \right)^{-1},
\end{equation}
one obtains 
\begin{equation}
    \frac{d}{d\epsilon} \big(
    \ln \hat{\rho}_{A}(\epsilon,t) \big)
    =
    \int_0^\infty d\beta\,
    \hat{G}_\epsilon(\beta,t)\,
    \Wop(t)\,
    \hat{G}_\epsilon(\beta,t).
    \label{eq:apdx_dlog}
\end{equation}
We now look at contribution from each order below:
\subsection{Absence of the first-order contribution}

Differentiating the entropy $S\!\left(\hat{\rho}_{A}(\epsilon,t)\right)$ gives
\begin{align}
\frac{d}{d\epsilon}
S\!\left(\hat{\rho}_{A}(\epsilon,t)\right)
&=
-\Tr\!\left[
    \Wop(t)\ln \hat{\rho}_{A}(\epsilon,t)
\right]
\nonumber\\
&\quad
-\Tr\!\left[
    \hat{\rho}_{A}(\epsilon,t)
    \frac{d}{d\epsilon}
    \ln \hat{\rho}_{A}(\epsilon,t)
\right].
\label{eq:apdx_first_derivative_entropy}
\end{align}
The second trace term in Eq.~\eqref{eq:apdx_first_derivative_entropy} vanishes which can be seen as follows.  We use  Eq.~\eqref{eq:apdx_dlog} and write
\begin{align}
    g(\epsilon,t)
    &\equiv
    \Tr\!\left[
        \hat{\rho}_{A}(\epsilon,t)
        \frac{d}{d\epsilon}
        \ln \hat{\rho}_{A}(\epsilon,t)
    \right]
    \nonumber\\
    &=
    \int_0^\infty d\beta\,
    \Tr\!\left[
        \hat{\rho}_{A}(\epsilon,t)
        \hat{G}_\epsilon
        \Wop
        \hat{G}_\epsilon
    \right].
\end{align}
Since \(\hat{\rho}_{A}(\epsilon,t)\) commutes with any function of itself, this can
be evaluated in the eigenbasis
\(\hat{\rho}_{A}(\epsilon,t)=\sum_i p_i(\epsilon,t)
\ket{i(\epsilon)}\bra{i(\epsilon)}\):
\begin{align}
    g(\epsilon,t)
    &=
    \sum_i
    \bra{i(\epsilon)}\Wop(t)\ket{i(\epsilon)}
    \int_0^\infty d\beta\,
    \frac{p_i(\epsilon,t)}
    {\left[p_i(\epsilon,t)+\beta\right]^2}
    \nonumber\\
    &=
    \sum_i
    \bra{i(\epsilon)}\Wop(t)\ket{i(\epsilon)}
    =
    \Tr[\Wop(t)]
    =
    0.
\end{align}
Thus,
\begin{equation}
    \frac{d}{d\epsilon}S\big(\hat{\rho}_{A}(\epsilon,t)\big)
    =
    -
    \Tr\!\left[
        \Wop(t)\ln \hat{\rho}_{A}(\epsilon,t)
    \right].
    \label{eq:apdx_entropy_first_derivative_simple}
\end{equation}
Evaluating this at \(\epsilon=0\), we get
\begin{equation}
    S^{(1)}(t)
    =
    -
    \Tr\!\left[
        \Wop(t)\ln \rACA(t)
    \right]
    =
    0.
    \label{eq:apdx_S1_zero}
\end{equation}
The last equality follows because \(\rACA\), and hence \(\ln\rACA\), is
block diagonal in the subsystem symmetry-sector basis, while \(\Wop\) is
purely off-block diagonal. Hence, there is no linear contribution to
\(\EA(t)\).

\subsection{Second-order contribution}

Differentiating Eq.~\eqref{eq:apdx_entropy_first_derivative_simple} once
more and then setting \(\epsilon=0\), we obtain
\begin{equation}
    S^{(2)}(t)
    =
    -
    \int_0^\infty d\beta\,
    \Tr\!\left[
        \Wop \,
        \hat{G}_0(\beta,t) \,
        \Wop \,
        \hat{G}_0(\beta,t)
    \right],
    \label{eq:apdx_S2_operator}
\end{equation}
where
\begin{equation}
    \hat{G}_0(\beta,t)
    =
    \left(
        \rACA(t)+\beta\,\id
    \right)^{-1}.
\end{equation}

Let
\begin{equation}
    \rACA(t)
    =
    \sum_i p_i(t)
    \ket{\chi_i(t)}\bra{\chi_i(t)}
\end{equation}
be the spectral decomposition of the symmetry-dephased state, and define
\(
    w_{ij}(t)
    =
    \bra{\chi_i(t)}
    \Wop(t)
    \ket{\chi_j(t)}\).
Then Eq.~\eqref{eq:apdx_S2_operator} becomes
\begin{equation}
    S^{(2)}(t)
    =
    -
    \sum_{i,j}
    |w_{ij}(t)|^2
    \int_0^\infty d\beta\,
    \frac{1}
    {\left(p_i(t)+\beta\right)
     \left(p_j(t)+\beta\right)}.
\end{equation}
Using
\begin{equation}
    \int_0^\infty d\beta\,
    \frac{1}
    {(p_i+\beta)(p_j+\beta)}
    =
    \frac{\ln p_i-\ln p_j}{p_i-p_j},
    \label{eq:apdx_M2}
\end{equation}
%with the equal-eigenvalue limit
%\begin{equation}
    %\lim_{p_j\rightarrow p_i}
    %\frac{\ln p_i-\ln p_j}{p_i-p_j}
    %=
    %\frac{1}{p_i},
%\end{equation}
we obtain
\begin{equation}
    S^{(2)}(t)
    =
    -
    \sum_{i,j}
    |w_{ij}(t)|^2 \,
    \frac{\ln p_i(t)-\ln p_j(t)}
    {p_i(t)-p_j(t)}.
    \label{eq:apdx_S2_final}
\end{equation}
Therefore, the leading contribution to the entanglement asymmetry is
\begin{equation}
    \EA^{(2)}(t)
    =
    -\frac{1}{2}S^{(2)}
    =
    \frac{1}{2}
    \sum_{i,j}
    |w_{ij}(t)|^2 \,
    \frac{\ln p_i(t)-\ln p_j(t)}
    {p_i(t)-p_j(t)}.
    \label{eq:apdx_EA2_final}
\end{equation}
Since the kernel in Eq.~\eqref{eq:apdx_EA2_final} is positive for positive
\(p_i,p_j\), the leading contribution is positive and quadratic in the
off-sector coherence matrix \(\Wop(t)\). Thus, in the small symmetry-breaking
regime,
\begin{equation}
    \EA(t)
    =
    \EA^{(2)}(t)
    +
    \mathcal{O}\!\left(\|\Wop(t)\|^3\right).
    \label{eq:apdx_EA_leading_order}
\end{equation}

\subsection{General \(n\)-th order term}

The pattern is
\begin{equation}
\begin{split}
\frac{d^{\,n-1}}{d\epsilon^{\,n-1}}
\big(\ln \hat{\rho}_{A}(\epsilon,t)\big)
\bigg|_{\epsilon=0}
&=
(-1)^{n-2}(n-1)! \\
&\!\!\!\!\!\!\! \times
\int_0^\infty d\beta\,
\left(\hat{G}_0\Wop\right)^{n-1}
\hat{G}_0.
\end{split}
\label{eq:apdx_general_dlog}
\end{equation}
Using Eq.~\eqref{eq:apdx_entropy_first_derivative_simple}, this gives
\begin{equation}
    S^{(n)}(t)
    =
    (-1)^{n-1}(n-1)!
    \int_0^\infty d\beta\,
    \Tr\!\left[
        \Wop
        \left(\hat{G}_0\Wop\right)^{n-1}
        \hat{G}_0
    \right].
    \label{eq:apdx_Sn_operator}
\end{equation}
Equivalently, in the eigenbasis of \(\rACA(t)\),
\begin{equation}
\begin{split}
    S^{(n)}
    =&\,
    (-1)^{n-1}(n-1)!
    \sum_{i_1,\ldots,i_n}
    w_{i_1i_2}
    w_{i_2i_3}
    \cdots
    w_{i_ni_1}\\
    &\qquad \qquad \qquad \qquad \, \times  M_n(p_{i_1},\ldots,p_{i_n}),
    \label{eq:apdx_Sn_final}  
\end{split}
\end{equation}
where
\begin{equation}
    M_n(p_{i_1},\ldots,p_{i_n})
    =
    \int_0^\infty d\beta\,
    \prod_{\mu=1}^{n}
    \frac{1}{p_{i_\mu}+\beta}.
    \label{eq:apdx_Mn_def}
\end{equation}

Substituting Eq.~\eqref{eq:apdx_Sn_final} into
Eq.~\eqref{eq:apdx_EA_from_Sn} and setting \(\epsilon=1\), the perturbative
series for the entanglement asymmetry is
\begin{eqnarray}
%\begin{split}
\Delta S_A(t)
=&
\sum_{n\ge 2}^{\infty}\frac{(-1)^n}{n}
\sum_{i_1,\dots,i_n}
w_{i_1 i_2}(t)w_{i_2 i_3}(t)\cdots w_{i_n i_1}(t) \nonumber 
\\
& \times
M_n(p_{i_1},\dots,p_{i_n})(t).
%\end{split}
\end{eqnarray}

Thus, \(\EA(t)\) is a nonlinear functional of the off-sector coherence
operator \(\Wop(t)\), with coefficients determined by the instantaneous
spectrum of the symmetry-dephased state \(\rACA(t)\). In particular, the
absence of the first-order term shows that the leading response of
\(\EA(t)\) to weak symmetry breaking is quadratic in \(\Wop(t)\).

\bibliography{references}
%%%%%%%%%%%%%%%%%%%%%%%%%%%%%%%%%%%%%%%%%%%%%%%%%%%%%%%%%%%%%%%

\end{document}